# Structure and Morphology of Crystalline Metal Nanoparticles: Polyhedral Cubic Particles

## Klaus E. Hermann

Inorg. Chem. Dept., Fritz-Haber-Institut der Max-Planck-Gesellschaft,
Faradayweg 4-6, 14195 Berlin, Germany.

## Abstract

We examine nanoparticles (NPs) forming polyhedral sections of the ideal cubic lattice, simple (sc), body centered (bcc), and face centered (fcc) cubic, which are confined by facets characterized by densest and second densest $\{h\,k\,l\}$ monolayers of the lattice. Together with the constraint that the NPs exhibit the same point symmetry as the ideal cubic lattice, i.e. $O_h$, different types of generic NPs serve for the definition of general polyhedral cubic NPs. Their structural properties, such as shape, size, and surface facets, are discussed in analytical and numerical detail with visualization of characteristic examples. The geometric relationships of the model particles expressed by corresponding formulas and numerical tables can be used to estimate shapes and sizes of real compact metal nanoparticles observed by experiment.

# I. Introduction

Nanoparticles of many sizes, shapes, and composition have become the target of a large number of recent experimental and theoretical studies due to their exciting physical and chemical properties [a1]. Here we mention only important applications in medicine [a2] or in catalysis where catalytic metal nanoparticles have become ubiquitous [a3, a4].

Shapes of real metal nanoparticles (NP), observed by experiment, are determined by their individual atoms being exposed to different local environments depending on their location inside the particle. Metal atoms close to the particle surface experience fewer atom neighbors compared to those inside the particle volume. This influences their interatomic binding and affects the particle structure. The variation of atom environments in finite particles depends strongly on the particle size since the relative number of surface atoms compared with those of the inner particle core becomes smaller with increasing size. This suggests that deviations from a crystalline bulk structure with its equivalent atom centers arranged in three-dimensional periodicity become less important as the particle size increases.

In many cases, structural properties of metal NPs with only a few atoms do not reflect those of corresponding bulk crystals. Details depend on the specific cluster shape and elemental composition and there are no general guidelines as to interatomic distances or angles or as to symmetry. This is illustrated by theoretical studies on silver NPs up to $Ag_{12}$ [a5] where equilibrium structures are found to deviate substantially from those of sections of the face-centered cubic crystal describing bulk silver. Also these very small NPs offer different stable isomers with varying shape and structure [a5]. Larger compact metal NPs can also exhibit symmetry properties which are not compatible with those of bulk crystals. As examples, many alkaline earth and transition metal (Nickel, Cobalt) NPs in gas phase with up to 5000 atoms [a6, a7] are believed to form compact particles with icosahedral symmetry $I_h$ which cannot appear in perfect bulk crystals due to their 5-fold rotational axes. Their structure can be explained by the concept of polyhedral atom shell filling which yields preferred NP sizes connected with so-called magic numbers of atoms [a7, a8]. However, many metal NPs have been observed by experiment to exhibit cubic $O_h$ symmetry which can be associated with compact sections of cubic bulk crystal structures, both face- and body-centered cubic, or can be approximated accordingly [a9]. Here examples are Aluminum and Indium NPs in gas phase between 1000 and 10000 atoms [a6] which are suggested to form compact polyhedral particles of face-centered cubic structure where confining facets are described by sections of densest (low Miller index) monolayers referring to $\{h\ k\ l\}$ families. Amongst these cuboctahedral shapes enclosed by both triangular $\{1\ 1\ 1\}$ and square

{1 0 0}facets, have been discussed. The corresponding NPs represent a fairly good approximation to spherical NPs since the atoms of the different planar sections do not vary too much in their distance from the NP center. Also other high-symmetry structures representing compact sections of face-centered cubic bulk crystals have been proposed in the literature [a6, a7] as possible structures of compact metal NPs where we mention only octahedral NPs which will be discussed below. Finally, metal NPs of $O_h$ symmetry described by sections of body-centered cubic bulk crystals have been suggested in the literature [a6, a10]. Here theoretical structure studies can help to describe and classify ideal compact cubic nanoparticles and, thus, identify structural properties of many real metal nanoparticles.

In this work we examine ideal nanoparticles forming polyhedral sections of the ideal cubic lattice, simple (sc), body centered (bcc), and face centered (fcc) cubic. These particles are assumed to be confined by facets of densest and second densest monolayers described by Miller indexed {$h\,k\,l$} families, i.e. {1 0 0}, {1 1 0} for sc and bcc as well as {1 1 1}, {1 0 0} for fcc. Together with the constraint that the NPs exhibit the same point symmetry as the ideal cubic lattice, i.e. $O_h$, there are different types of generic NPs which serve for the definition of general polyhedral NPs. Their structural properties , such as shape, size, and surfaces, are discussed in analytical and numerical detail with visualization of characteristic examples. These results can also be used as a repository for structures of compact NPs with internal cubic lattice. All NP graphics and analyses have been obtained with the help of the Balsac software developed by the author [a11].

## II. Results and Discussion

## A. Nanoparticles with Simple Cubic (sc) Lattice Structure

The simple cubic (sc) lattice is defined by lattice vectors $\underline{R}_1$, $\underline{R}_2$, $\underline{R}_3$ in Cartesian coordinates according to

$$\underline{R}_1 = a(1,0,0) \, , \quad \underline{R}_2 = a(0,1,0) \, , \quad \underline{R}_3 = a(0,1,0) \tag{A.1}$$

where $a$ is the lattice constant. The three densest monolayer families of the sc lattice are described by six square shaped {1 0 0} (highest density), twelve rectangular {1 1 0}, and eight hexagonal {1 1 1} netplanes where distances between adjacent parallel netplanes are given by

$$d_{\{100\}} = a \, , \quad d_{\{110\}} = a/\sqrt{2} \, , \quad d_{\{111\}} = a/\sqrt{3} \tag{A.2}$$

The point symmetry of the sc lattice is characterized by $O_h$ with symmetry centers at all atom sites and at the void centers of each elementary cell.

Compact simple cubic nanoparticles (NPs) are confined by finite sections of monolayers (facets) whose structure is described by different netplanes ($h\,k\,l$). If they exhibit central $O_h$ symmetry and show an ($h\,k\,l$) oriented facet they must must also include all other symmetry related facets characterized by orientations of the complete $\{h\,k\,l\}$ family. Thus, general sc NPs of $O_h$ symmetry are determined by facets whose orientation can be defined by those of different $\{h\,k\,l\}$ families. (As an example, we mention the {1 0 0} family with its six netplane orientations ($\pm 1\,0\,0$), ($0\,\pm 1\,0$), ($0\,0\,\pm 1$).) Further, according to the symmetry of the sc host lattice possible NP centers can only be atom or $O_h$ symmetry void sites of the lattice. This will be discussed at different levels of complication in the following.

### A.1. Generic sc Nanoparticles

First, we consider generic sc nanoparticles (NPs) of $O_h$ symmetry which are confined by facets with orientations of only one netplane family $\{h\,k\,l\}$. This allows to distinguish between different generic NP types where we focus on those confined by {1 0 0} or {1 1 0} facets which correspond to the densest monolayers of the sc lattice and offer the flattest NP facets.

(a) **Generic cubic** sc NPs, denoted **sc[$N$, 0)** (the notation of sc NPs, in particular the bracketing, will be explained in Secs. A.2, 3), contain an **atom** at their symmetry center for **even** $N$ and a **void** for **odd** $N$. They are confined by all six {1 0 0} monolayers with distances $D_{\{100\}} = N\, d_{\{100\}}$ between parallel monolayers (polyhedral NP diameters). This yields six square shaped {1 0 0} facets with ($N + 1$) edge atoms each and eight polyhedral atom corners, see Fig. A.1.

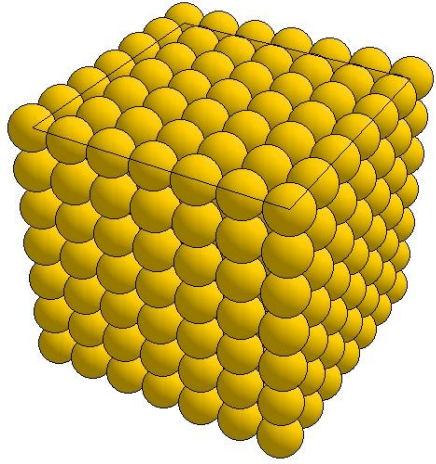

**Figure A.1.** Atom ball models of a generic cubic sc NP, sc[6, 0) with an atom at its center. The black lines sketch the square {1 0 0} facet shapes.

(b) **Generic rhombic** sc NPs are confined by all twelve {1 1 0} monolayers with distances $D_{\{110\}} = 2M\, d_{\{110\}}$ between parallel monolayers (polyhedral NP diameters). Here we distinguish two NP types.

**Generic rhombic A** sc NPs, denoted **sc(0, M]**, contain an **atom** at their symmetry center and are confined by rhombic facets of all twelve {1 1 0} monolayers, see Figs. A.2. For even $M$, Fig. A.2a, the {1 1 0} rhombi contain ($M + 1$) parallel nearest neighbor atom rows of increasing/decreasing length in the sequence 0 (atom), $2a$, …, $M\,a$, …, $2a$, 0 (atom). For odd $M$, Fig. A.2b, the lengths increase/decrease in the sequence 0 (atom), $2a$, …, $(M-1)a$, $(M-1)a$, …, $2a$, 0 (atom). This yields a capping of the rhombi at their near corners resulting in eight additional triangular {1 1 1} microfacets at the sc NP surface. Altogether, the generic rhombic A sc NPs are described as rhombic dodecahedra offering 14 polyhedral atom corners for even $M$ and reminding of the shape of Wigner-Seitz cells of the face centered cubic (fcc) crystal lattice [a12]. For odd $M$, the dodecahedra are capped at eight corners which leads to 30 polyhedral atom corners.

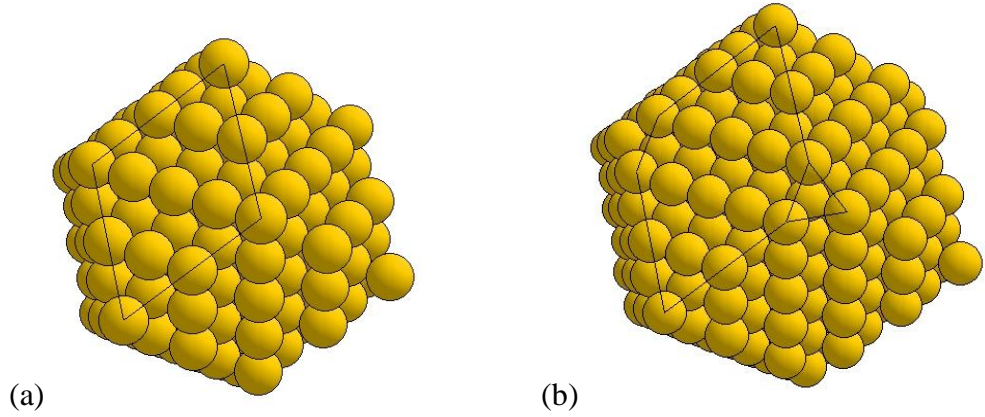

(a)          (b)

**Figure A.2.** Atom ball models of generic rhombic A sc NPs with an atom at their symmetry centers, (a) sc(0, 4] and (b) sc(0, 5]. The black lines sketch the (capped) rhombic {1 1 0} and triangular {1 1 1} facet shapes.

**Generic rhombic B** fcc NPs, denoted **sc(1, *M*]**, contain a **void** at their symmetry center and are also confined by rhombic facets of all twelve {1 1 0} monolayers, see Figs. A.3. For even *M*, Fig. A.3a, the {1 1 0} rhombi contain *M* parallel nearest neighbor atom rows of increasing/decreasing length in the sequence $a$, $3a$, …, $(M-1)a$, $(M-1)a$, …, $3a$, $a$. This yields a capping of the rhombi at their far corners resulting in six additional square {1 0 0} microfacets at the sc NP surface. For odd *M*, Fig. A.3b, the lengths increase/decrease in the sequence $a$, $3a$, …, $M a$, …, $3a$, $a$ which yields a capping of the rhombi at all four corners resulting in eight additional triangular {1 1 1} microfacets. Altogether, the generic rhombic B sc NPs are described by capped rhombic dodecahedra offering 48 (32) polyhedral atom corners for even (odd) *M* and.

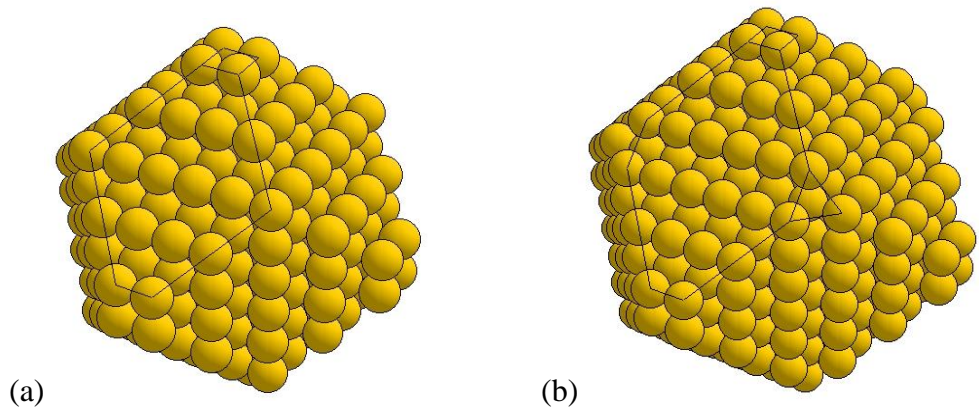

(a)          (b)

**Figure A.3.** Atom ball models of generic rhombic B sc NPs with a void at their symmetry centers, (a) sc(1, 5] and (b) sc(1, 6]. The black lines sketch the (capped) rhombic {1 1 0}, the triangular {1 1 1}, and the square {1 0 0} facet shapes.

7## A.2. Non-generic sc Nanoparticles

Non-generic sc nanoparticles of $O_h$ symmetry are confined by facets with orientations of several netplane families $\{h_i\, k_i\, l_i\}$. This can be considered as combining the confinements of the corresponding generic NPs, $NP_i$, determined by $\{h_i\, k_i\, l_i\}$ with suitable polyhedral diameters, sharing their symmetry center and type (atom or void). As an example we discuss non-generic sc nanoparticles which combine two generic nanoparticles $NP_1$, $NP_2$ of the types detailed in Sec. A.1. Thus, the resulting sc NP contains only atoms inside both partner NPs. Here three possible scenarios can be distinguished, nanoparticle $NP_2$ contains all atoms of $NP_1$ (yielding generic $NP_1$), $NP_1$ contains all atoms of $NP_2$ (yielding generic $NP_2$), or $NP_1$ and $NP_2$ share only parts of their atoms (intersecting, which yields true non-generic NPs). These scenarios are determined by the choice of the two netplane families $\{h\, k\, l\}$ and by relationships between the corresponding polyhedral NP diameters $D_{\{hkl\}}$ defined by $N$ and $M$. In the following, we restrict ourselves to non-generic sc NPs defined by pairs of generic cubic and rhombic NPs.

Non-generic sc NPs with an **atom** at their symmetry center combine a generic cubic NP, $sc[N, 0]$ with even $N$, and a rhombic A NP, $sc(0, M]$ where polyhedral diameters amount to

$$D_{\{100\}} = N\, d_{\{100\}} = N\, a, \quad N \text{ even}, \qquad D_{\{110\}} = 2M\, d_{\{110\}} = 2M\, a/\sqrt{2} \qquad (A.3)$$

Thus, the smallest rhombic A NP which surrounds the cubic NP is defined with (A.2) by

$$D_{\{110\}} = \sqrt{2}\, D_{\{100\}} \quad \text{and hence} \quad M = N \qquad (A.4)$$

On the other hand, the smallest cubic NP which surrounds the rhombic A NP is defined with (A.2) by

$$D_{\{100\}} = \sqrt{2}\, D_{\{110\}} \quad \text{and hence} \quad N = 2M \qquad (A.5)$$

As a consequence, non-generic sc NPs with an atom at their symmetry center combining generic cubic and rhombic A NPs, $NP_1$, $NP_2$, yield the choices shown in Table A.1.

| $NP_1$ / $NP_2$ | $NP_1$ in $NP_2$ | $NP_2$ in $NP_1$ |
|---|---|---|
| Cubic / Rhombic A<br>$sc[N, 0]$ / $sc(0, M]$, $N$ even | $M \geq N$ | $N \geq 2M$ |

**Table A.1.** Constraints of structure parameters $N$, $M$ for pairs of sc NPs, $NP_1$ and $NP_2$, sharing an atom as their symmetry center.

Non-generic sc NPs with a **void** at their symmetry center combine a generic cubic NP, sc[$N$, 0) with odd $N$, and a rhombic B NP, sc(1, $M$] where polyhedral diameters amount to

$$D_{\{100\}} = N\, d_{\{100\}} = N\, a, \quad N \text{ odd}, \quad D_{\{110\}} = 2M\, d_{\{110\}} = 2M\, a/\sqrt{2} \tag{A.6}$$

which also leads to relation (A.4) for the smallest rhombic B NP surrounding the cubic NP. Further, the smallest cubic NP which surrounds the rhombic B NP is given by

$$(D_{\{100\}} + d_{\{100\}}) = \sqrt{2}\, D_{\{110\}} \quad \text{and hence} \quad N = 2M - 1 \tag{A.7}$$

Therefore, non-generic sc NPs with a void at their symmetry center combining generic cubic and rhombic B NPs, $NP_1$, $NP_2$, yield the three choices shown in Table A.2.

| $NP_1$ / $NP_2$ | $NP_1$ in $NP_2$ | $NP_2$ in $NP_1$ |
|---|---|---|
| Cubic / Rhombic B<br>sc[$N$, 0) / sc(1, $M$], $N$ odd | $M \geq N$ | $N \geq 2M-1$ |

**Table A.2.** Constraints of structure parameters $N$, $M$ for pairs of sc NPs, $NP_1$ and $NP_2$, sharing a void as their symmetry center.

Tables A.1, 2 show in particular that true non-generic sc NPs, defined by pairs of generic cubic NPs, sc[$N$, 0), and rhombic NPs, sc($q$, $M$], $q = 0, 1$, refer to structure parameters $N$, $M$ which bracket each other according to

$$M < N < 2M, \quad N/2 < M < N \qquad N \text{ even, atom centered} \tag{A.8a}$$

$$M < N < 2M - 1, \quad (N+1)/2 < M < N \qquad N \text{ odd, void centered} \tag{A.8b}$$

These non-generic sc NPs may be called **cubo-rhombic** and denoted as **sc[$N$, $M$]** combining a generic cubic sc[$N$, 0) (first index) with a rhombic NP (second index) where the latter is rhombic A, sc(0, $M$], for even $N$ and rhombic B, sc(1, $M$], for odd $N$. The nomenclature together with the results of Tables A.1, 2 suggests an alternative notation for generic NPs where we may formally write

generic cubic  sc[$N$, 0) = sc[$N$, $M$] ,  $N \leq M$  (A.9a)

generic rhombic A  sc(0, $M$] = sc[$N$, $M$] ,  $N \geq 2M$  (A.9b)

generic rhombic B  sc(1, $M$] = sc[$N$, $M$] ,  $N \geq 2M - 1$  (A.9c)

## A.3. Cubo-rhombic sc Nanoparticles

In the following we will discuss the detailed structure of true cubo-rhombic NPs sc[$N$, $M$] determined by corresponding structure parameters $N$, $M$ and confined by the two densest monolayer families {1 0 0} and {1 1 0}. These NPs contain an **atom** at there symmetry center for **even** $N$ while their symmetry center is **void** for **odd N**.

First, we consider **atom centered** sc NPs. Starting from a generic cubic NP sc[$N$, 0) ($N$ even) with a polyhedral NP diameter $D_{\{100\}} = N\,d_{\{100\}}$ the smallest enclosing rhombic A NP sc(0, $M$] has a polyhedral NP diameter $D_{\{110\}} = 2M\,d_{\{110\}}$ with $M = N$ according to Tables A.1, 2. Shrinking the rhombic sc NP, i.e. for smaller $M$ values given by $M = N - m$, $m \geq 0$, the cubic and rhombic sc NPs intersect yielding a **cubo-rhombic A** NP sc[$N$, $M = N - m$]. In the following we define this NP by **sc[$N$, $m$)** which is compatible with the initial definition of a generic cubic sc NP denoted sc[$N$, 0) in Sec. A.1. The constraints on $M$ given in Table A.1 can be expressed by equivalent constraints on $m$ yielding

$$N/2 \leq M \leq N \;, \qquad 0 \leq m \leq N/2 \qquad \text{cubo-rhombic A} \qquad (A.10)$$

where $m$ values larger than $N/2$ result in generic rhombic A sc NPs. The building scheme corresponds to the generic cubic NP sc[$N$, 0) being truncated at its twelve edges by removing $m$ {1 1 0} facet layers each.

On the other hand, starting from a generic rhombic A NP sc(0, $M$] with a polyhedral NP diameter $D_{\{110\}} = 2M\,d_{\{110\}}$ the smallest enclosing cubic sc NP has a polyhedral NP diameter $D_{\{100\}} = N\,d_{\{100\}}$ with $N = 2M$ according to Table A.1. Shrinking the cubic sc NP, i.e. for smaller $N$ values given by $N = 2M - n$ with $n = 0, 2, 4 \ldots$ the cubic and rhombic NPs intersect yielding also a **cubo-rhombic A** NP sc[$N = 2M - n$, $M$]. In the following we define this NP by **sc($n$, $M$]** with even $n$ which is compatible with the initial definition of a generic rhombic A sc NP denoted sc(0, $M$] in Sec. A.1. The constraints on $N$ given in Table A.1 can be expressed by equivalent constraints on $n$ yielding

$$M \leq N \leq 2M \;, \qquad 0 \leq n \leq M \;, \quad n \text{ even} \qquad \text{cubo-rhombic A} \qquad (A.11)$$

where $n$ values larger than $M$ result in generic cubic sc NPs. The building scheme corresponds to the generic rhombic A NP sc(0, $M$] being truncated at its six 4-fold symmetry corners by removing $n/2$ {1 0 0} facet layers each.

The two building schemes must yield the same cubo-rhombic A NP sc[$N$, $M$] where the two structure parameters $N$, $M$ quantify the polyhedral NP diameters $D_{\{100\}}$ and $D_{\{110\}}$ according to

$$D_{\{100\}} = N\,d_{\{100\}} \;, \qquad D_{\{110\}} = 2M\,d_{\{110\}} \qquad (A.12)$$

with the basic netplane distances $d_{\{100\}}$ and $d_{\{110\}}$ given by (A.2). Further, the relations

$$M = N - m \;, \qquad N = 2M - n \;, \quad n \text{ even} \qquad (A.13a)$$

from the two building schemes can be converted to

$$n = 2M - N, \qquad m = N - M \qquad (A.13b)$$

$$N = n + 2m, \qquad M = n + m \qquad (A.13c)$$

Next, we consider **void centered** sc NPs whose treatment is completely analogous to the atom centered case. Starting from a generic cubic NP sc[$N$, 0] ($N$ odd) and its smallest enclosing rhombic B NP sc(1, $M$] with $M = N$, shrinking the rhombic sc NP yields a **cubo-rhombic B** NP sc[$N$, $M = N - m$] which can be defined by **sc[$N$, $m$)**. The constraints on $M$ given in Table A.2 can be expressed by equivalent constraints on $m$ yielding

$$(N+1)/2 \leq M \leq N, \qquad 0 \leq m \leq (N-1)/2 \qquad \text{cubo-rhombic B} \qquad (A.14)$$

where $m$ values larger than $(N-1)/2$ result in generic rhombic B sc NPs. The building scheme corresponds also to the generic cubic NP sc[$N$, 0] being truncated at its twelve edges by removing $m$ {1 1 0} facet layers each.

On the other hand, starting from a generic rhombic B NP sc(1, $M$] and its smallest enclosing cubic sc NP sc [$N$, 0] with $N = 2M - 1$, shrinking the cubic sc NP yields also a **cubo-rhombic B** NP sc[$N = 2M - n$, $M$] which can be defined by **sc[$n$, $M$)** with odd $n$. This includes the initial definition of a generic rhombic B sc NP denoted sc(1, $M$]. The constraints on $N$ given in Table A.2 can be expressed by equivalent constraints on $n$ yielding

$$M \leq N \leq 2M-1, \qquad 1 \leq n \leq M, \quad n \text{ odd} \qquad \text{cubo-rhombic B} \qquad (A.15)$$

where $n$ values larger than $M$ result in generic cubic sc NPs. The building scheme corresponds to the generic rhombic B NP sc(1, $M$], being truncated at its six 4-fold symmetry corners by removing $(n - 1)/2$ {1 0 0} facet layers each.

The two building schemes must yield the same cubo-rhombic B NP sc[$N$, $M$] resulting in

$$M = N - m, \qquad N = 2M - n, \quad n \text{ odd} \qquad (A.16)$$

and hence to relations (A.13b), (A.13c) derived above.

Altogether, relations (A.13) suggest an alternative set of structure parameters $n$, $m$ which characterize the deviation of the cubo-rhombic NP sc[$N$, $M$] from its generic envelope NPs, cubic sc[$N$, 0] and rhombic A sc(0, $M$] or rhombic B sc(1, $M$]. Thus, **cubo-rhombic A** (atom centered) or **B** (void centered) NPs can also be denoted **sc($n$, $m$)** with even or odd $n$, respectively.

The structure parameters $n$, $m$ are also connected with geometric properties of the NP facets, in particular, with characteristic lengths of the facet edges. The cubo-rhombic sc NPs, sc[$N$, $M$] ≡

sc($n, m$), are confined by all six {1 0 0} facets of square shape with $(n + 1) \times (n + 1)$ atoms each and by all twelve {1 1 0} facets, see Figs. A.4. For even $m$, Fig. A.4a, the {1 1 0} facets are hexagonal and contain $m + 1$ parallel nearest neighbor atom rows of increasing/decreasing length in the sequence $n\,a$, $(n+2)a$, …, $(n + m)\,a$, …, $(n+2)a$, $n\,a$. For odd $m$, Fig. A.4b, the {1 1 0} facets are capped hexagonal (octagonal) and the lengths of the $m + 1$ atom rows increase/decrease in the sequence $n\,a$, $(n+2)a$, …, $(n + m -1)a$, $(n + m -1)a$, …, $(n+2)a$, $n\,a$. This yields eight additional triangular {1 1 1} microfacets shown in Fig A.5b. Altogether, the sc NPs are described by cubo-rhombic polyhedra offering 32 ($n > 0$, $m > 0$, $m$ even) or 48 ($n > 0$, $m > 0$, $m$ odd) corners with true cubic ($n > 0$, $m = 0$, eight corners) and true rhombic A ($n = 0$, $m > 0$, 14 corners) being special cases.

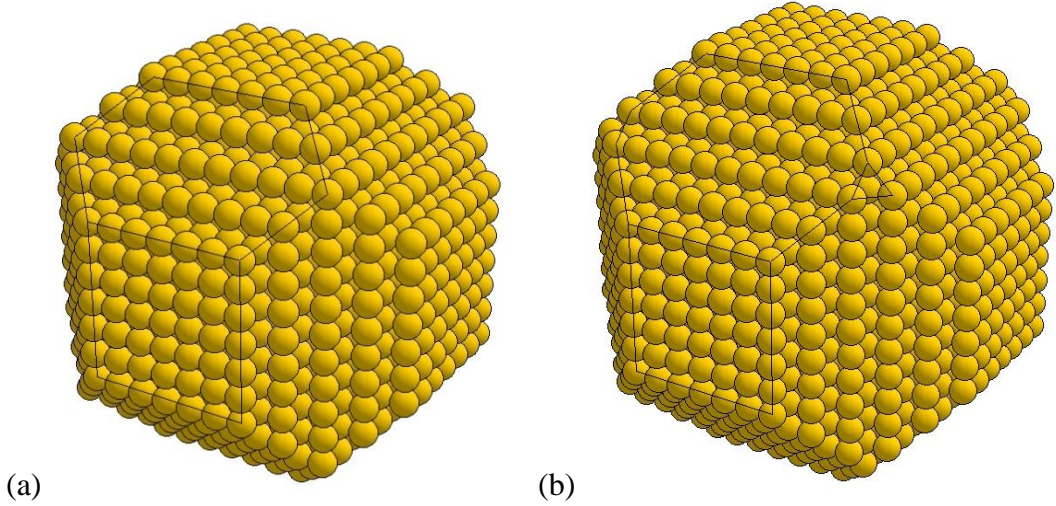

(a)  (b)

**Figure A.4.** Atom ball models of atom centered cubo-rhombic A sc NPs, (a) sc[14, 10] ≡ sc(6, 4) and (b) sc[16, 11] ≡ sc(6, 5). The black lines sketch the square {1 0 0}, (capped) hexagonal {1 1 0} and triangular {1 1 1} facet shapes.

The confinement of a cubo-rhombic A sc[$N, M$] ≡ sc($n, m$) NP with an atom at its symmetry center ($N$, $n$ even) by its generic envelope sc NPs, cubic sc[$N, 0$) and rhombic A sc[$0, M$] with $N, M$ defined by A.13, is illustrated in Fig. A.5 for sc[14, 10] ≡ sc(6, 4) with sc[14,0) and sc(0, 10] forming the envelope sc NPs.

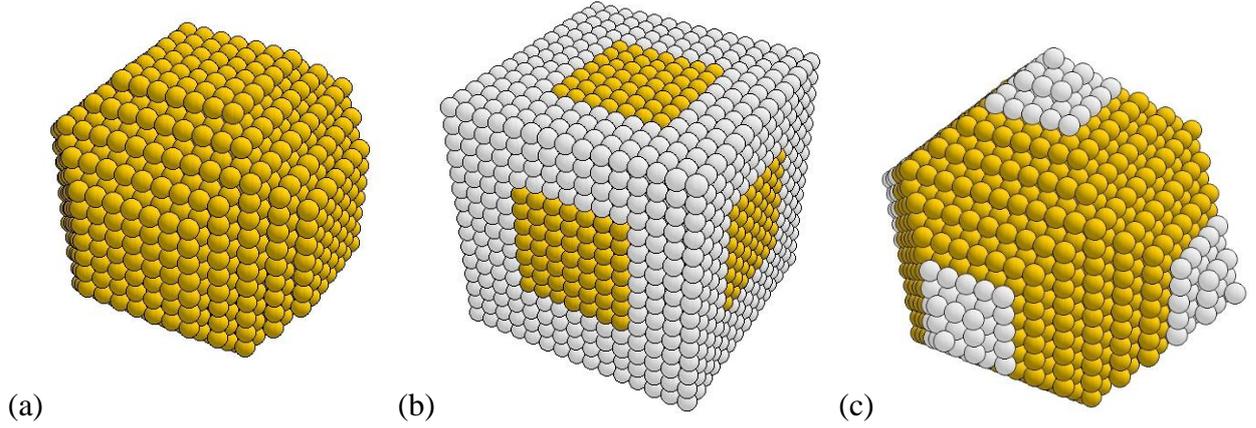

**Figure A.5.** Atom ball models of a cubo-rhombic A sc[14, 10] ≡ sc(6, 4) NP with its confining generic cubic and rhombic A NPs. (a) Initial sc(6, 4) shown by yellow balls, (b) sc(6, 4) inside generic sc[14, 0], and (c) sc(6, 4) inside generic sc(0, 10). The atoms outside sc(6, 4) are shown by white balls.

Further, the confinement of a cubo-rhombic B sc[$N$, $M$] ≡ sc($n$, $m$) NP with a void at its symmetry center ($N$, $n$ odd) by its cubic sc[$N$, 0] and rhombic B sc(1, $M$) envelope sc NPs is shown in Fig. 6 for sc[13, 9] ≡ sc(5, 4) with sc[13,0] and sc(1, 9) forming the envelopes.

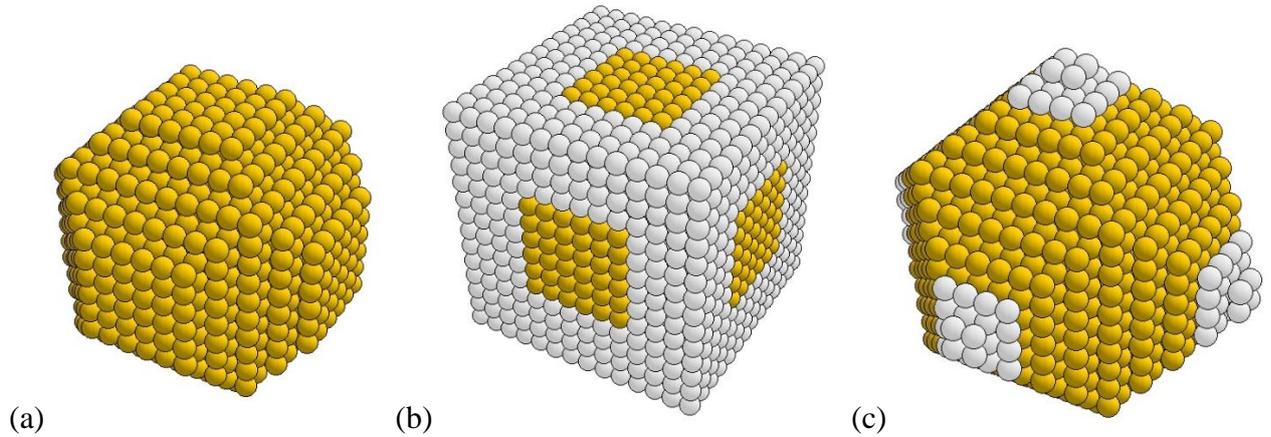

**Figure A.6.** Atom ball models of a cubo-rhombic B sc[13, 9] ≡ sc(5, 4) NP with its confining generic cubic and rhombic B NPs. (a) Initial sc(5, 4) shown by yellow balls, (b) sc(5, 4) inside generic sc[13, 0], and (c) sc(5, 4) inside generic sc(1, 9). The atoms outside sc(5, 4) are shown by white balls.

Altogether, the equivalence of structure parameters $n$, $m$ and $N$, $M$ according to (A.13) suggests alternative notations of cubo-rhombic sc NPs introduced earlier. Apart from a notation by sc[$N$, $M$], meaningful alternatives are sc($n$, $m$), sc[$N$, $m$], and sc($n$, $M$] where the different bracketing denotes the different meanings of the two parameters in the notation. Table A.3 lists types and notations of all cubo-rhombic sc NPs where type A NPs (atom centered) refer to even $n$, $N$ values while for type B NPs (void centered) $n$, $N$ values are odd.

| Type | sc[$N$, $M$] | sc($n$, $m$) | sc[$N$, $m$] | sc($n$, $M$) |
|---|---|---|---|---|
| Cubo-rhombic A | $M < N < 2M$ | $n > 0, m > 0$ | $0 < m < N/2$ | $0 < n < M$ |
| Cubo-rhombic B | $M < N < 2M-1$ | $n > 1, m > 0$ | $0 < m < (N-1)/2$ | $1 < n < M$ |
| Generic cubic | $N = M$ | $n > 0, m = 0$ | $m = 0$ | $n = M$ |
| Generic rhombic A | $N = 2M$ | $n = 0, m > 0$ | $m = N/2$ | $n = 0$ |
| Generic rhombic B | $N = 2M-1$ | $n = 1, m > 0$ | $m = (N-1)/2$ | $n = 1$ |
| Elementary cube (B) | $N = M = 1$ | $n = 1, m = 0$ | $N = 1, m = 0$ | $n = M = 1$ |
| Atom (A) | $N = M = 0$ | $n = m = 0$ | $N = m = 0$ | $n = M = 0$ |

**Table A.3.** Types and notations of all intersecting and generic sc NPs.

## A.4. Properties of Cubo-rhombic sc Nanoparticles

Structural properties characterizing cubo-rhombic sc nanoparticles can be obtained by geometric consideration and simple algebra where we quote only results of the most important quantities. Structure parameters $n$, $m$, $N$, $M$, according to (A.13), i.e.

$$n = 2M - N, \qquad m = N - M \qquad (A.13b)$$
$$N = n + 2m, \qquad M = n + m \qquad (A.13c)$$

with constraints listed in Tables A.1 - 3 will be used interchangeably and in mixed combinations to simplify formal expressions. Note that according to (A.13) structure parameters $n$ and $N$ will always be both even or both odd numbers. Results which are valid only for atom centered, type A, or for void centered, type B, sc NPs will be denoted accordingly.

The number of atoms forming an sc($n$, $m$) NP are given by the **volume count** denoted $N_{vol}(n, m)$ where simple algebra yields for

**- $m$ even** ($N$, $M$ both even or both odd)

$$N_{vol}(n, m) = (n+m+1)^3 + (n+m+1)(n+m)(n+m-1) - n(n^2-1)$$
$$= (M+1)^3 + M(M^2-1) - (2M-N)\left[(2M-N)^2 - 1\right] \qquad (A.17a)$$

- **m odd** (*N* even, *M* odd or *N* odd, *M* even)

$$N_{vol}(n, m) = (n+m+1)^3 + (n+m+1)(n+m)(n+m-1) - n(n^2-1) - 1$$
$$= (M+1)^3 + M(M^2-1) - (2M-N)\left[(2M-N)^2 - 1\right] - 1 \quad \text{(A.17b)}$$

with **special cases**

| | | | |
|---|---|---|---|
| Generic cubic | $N_{vol}(n, 0) = (n+1)^3 = (N+1)^3$ | $N = n$ | (A.18) |

Generic rhombic A
$$N_{vol}(0, m) = m(2m^2 + 3m + 2) + 1 \quad \text{(A.19a)}$$
$$= M(2M^2 + 3M + 2) + 1 \quad M = m \text{ even}$$
$$N_{vol}(0, m) = m(2m^2 + 3m + 2) \quad \text{(A.19b)}$$
$$= M(2M^2 + 3M + 2) \quad M = m \text{ odd}$$

Generic rhombic B
$$N_{vol}(1, m) = (m+1)(2m^2 + 7m + 7) + 1 \quad \text{(A.20a)}$$
$$= M(2M^2 + 3M + 2) + 1 \quad M = m + 1 \text{ odd}$$
$$N_{vol}(1, m) = (m+1)(2m^2 + 7m + 7) \quad \text{(A.20b)}$$
$$= M(2M^2 + 3M + 2) \quad M = m + 1 \text{ even}$$

The number of atoms on the outermost facets of an sc(*n, m*) NP (forming the outer polyhedral shell confining the NP) are given by the **facet count** denoted $N_{shell}(n, m)$ where simple algebra yields for

- **m even** (*N, M* both even or both odd)

$$N_{shell}(n, m) = 6(m+n)^2 + 2 = 6M^2 + 2 \quad \text{(A.21a)}$$

- **m odd** (*N* even, *M* odd or *N* odd, *M* even)

$$N_{shell}(n, m) = 6(m+n)^2 = 6M^2 \quad \text{(A.21b)}$$

with **special cases**

| | | | |
|---|---|---|---|
| Generic cubic | $N_{shell}(n, 0) = 6n^2 + 2 = 6N^2 + 2$ | $N = n$ | (A.22) |
| Generic rhombic A | $N_{shell}(0, m) = 6m^2 + 2 = 6M^2 + 2$ | $M = m$ even | (A.23a) |
| | $N_{shell}(0, m) = 6m^2 = 6M^2$ | $M = m$ odd | (A.23b) |
| Generic rhombic B | $N_{shell}(1, m) = 6(m+1)^2 + 2 = 6M^2 + 2$ | $M = m + 1$ odd | (A.24a) |
| | $N_{shell}(1, m) = 6(m+1)^2 = 6M^2$ | $M = m + 1$ even | (A.24b) |

The distances between the center of an sc($n$, $m$) NP and its outer facet midpoints (**facet distances**) are given by $R_{\{hkl\}}(n, m)$, $\{h\,k\,l\} = \{1\,0\,0\}$, $\{1\,1\,0\}$, and $\{1\,1\,1\}$, which are connected with the polyhedral NP diameters $D_{\{hkl\}}$ according to

$$R_{\{hkl\}}(n,m) = \frac{1}{2}D_{\{hkl\}} \tag{A.25}$$

Thus, with (A.2), (A.3), (A.6), (A.7), (A.13) and $a$ denoting the lattice constant of the sc lattice we obtain

$$R_{\{100\}}(n,m) = \frac{1}{2}N d_{\{100\}} = \frac{a}{2}(n+2m) \tag{A.26}$$

$$R_{\{110\}}(n,m) = \frac{1}{2}(2M)d_{\{110\}} = \frac{a}{\sqrt{2}}(n+m) \tag{A.27}$$

$$R_{\{111\}}(n,m) = \frac{1}{2}(3M)d_{\{111\}} = \frac{\sqrt{3}}{2}a(n+m) \qquad m \text{ even }^+ \tag{A.28a}$$

$$= \frac{1}{2}(3M-1)d_{\{111\}} = \frac{\sqrt{3}}{2}a\left(n+m-\frac{1}{3}\right) \qquad m \text{ odd} \tag{A.28b}$$

+ Note that for even $m$ all $\{1\,1\,1\}$ facets reduce to one atom. Thus, distances $R_{\{111\}}(n, m)$ refer to corner rather than facet distances.

The distances between the center of an sc($n$, $m$) NP and its two different types of corners (**corner distances**) are given by $R_{c1}(n, m)$ (referring to each of six ($n = 0$) or 24 ($n > 0$) corners at $\{1\,0\,0\}$ facets) and by $R_{c2}(n, m)$ (referring to each of eight ($m$ even) or 24 ($m$ odd) corners at $\{1\,1\,0\}$ (and eight possible $\{1\,1\,1\}$) facets for odd $m$) where with (A.7) and $a$ denoting the lattice constant of the sc lattice

$$R_{c1}(n,m) = \frac{a}{2}\sqrt{(n+2m)^2 + 2n^2} = \frac{a}{2}\sqrt{N^2 + 2(2M-N)^2} \tag{A.29}$$

$$R_{c2}(n,m) = \frac{\sqrt{3}}{2}a(n+m) = \frac{\sqrt{3}}{2}aM \qquad m \text{ even} \tag{A.30a}$$

$$= \frac{a}{2}\sqrt{2(n+m-1)^2 + (n+m+1)^2} = \frac{a}{2}\sqrt{3(M-1)^2 + 4M} \qquad m \text{ odd} \tag{A.30b}$$

The areas of each of the $\{1\,0\,0\}$, $\{1\,1\,0\}$, and $\{1\,1\,1\}$ facets of an sc($n$, $m$) NP, measured by corresponding corner atoms (**facet areas**), are given by $A_{\{100\}}(n, m)$ (referring to each of six facets), $A_{\{110\}}(n, m)$ (referring to each of twelve facets), and $A_{\{111\}}(n, m)$ (referring to each of six microfacets, for odd $m$ only) where, with (A.6), (A.7) and $a$ denoting the lattice constant of the sc lattice

$$A_{\{100\}}(n,m) = a^2 n^2 = a^2 (2M - N)^2 \tag{A.31}$$

$$A_{\{110\}}(n,m) = \frac{1}{\sqrt{2}} a^2 m(2n + m)$$

$$= \frac{1}{\sqrt{2}} a^2 (N - M)(3M - N) \qquad m \text{ even} \tag{A.32a}$$

$$A_{\{110\}}(n,m) = \frac{1}{\sqrt{2}} a^2 \left[ m(2n + m) - 1 \right]$$

$$= \frac{1}{\sqrt{2}} a^2 \left[ (N - M)(3M - N) - 1 \right] \qquad m \text{ odd} \tag{A.32b}$$

$$A_{\{111\}}(n,m) = \frac{\sqrt{3}}{2} a^2 \tag{A33}$$

## B. Nanoparticles with Body Centered Cubic (bcc) Lattice Structure

The body centered cubic (bcc) lattice can be defined as a non-primitive simple cubic lattice by lattice vectors $\underline{R}_1$, $\underline{R}_2$, $\underline{R}_3$ in Cartesian coordinates together with two lattice basis vectors $\underline{r}_1$, $\underline{r}_2$ according to

$$\underline{R}_1 = a(1,0,0) \ , \quad \underline{R}_2 = a(0,1,0) \ , \quad \underline{R}_3 = a(0,1,0) \tag{B.1a}$$

$$\underline{r}_1 = a(0,0,0) \ , \quad \underline{r}_2 = \frac{a}{2}(1,1,1) \tag{B.1b}$$

where $a$ is the lattice constant. The two densest monolayer families of the bcc lattice are described by six square shaped {1 0 0}, twelve rectangular {1 1 0} (highest density), and eight hexagonal {1 1 1} netplanes where distances between adjacent parallel netplanes are given by

$$d_{\{100\}} = a/2 \ , \quad d_{\{110\}} = a/\sqrt{2} \ , \quad d_{\{111\}} = a/(2\sqrt{3}) \tag{B.2}$$

The point symmetry of the bcc lattice is characterized by $O_h$ with symmetry centers at all atom sites.

Compact body centered cubic nanoparticles (NPs) are confined by finite sections of monolayers (facets) whose structure is described by different netplanes (*h k l*). If they exhibit central $O_h$ symmetry and show an (*h k l*) oriented facet they must must also include all other symmetry related facets characterized by orientations of the complete {*h k l*} family. Thus, general bcc NPs of $O_h$ symmetry are determined by facets whose orientation can be defined by those of different {*h k l*} families. (As an example, we mention the {1 0 0} family with its six netplane orientations (±1 0 0), (0 ±1 0), (0 0 ±1).) Further, according to the symmetry of the bcc host lattice possible NP centers can only be atom sites of the lattice. This will be discussed at different levels of complication in the following.

### B.1. Generic bcc Nanoparticles

First, we consider generic bcc nanoparticles (NPs) of $O_h$ symmetry which are confined by facets with orientations of only one netplane family {*h k l*}. This allows to distinguish between different generic NP types where we focus on those described by {1 0 0} or {1 1 0} facets which correspond to the densest monolayers of the bcc lattice and offer the flattest NP facets. The bcc NPs contain always an **atom** at their symmetry center.

(a) **Generic cubic** bcc NPs, denoted **bcc[*N*, 0)** (the notation of bcc NPs, in particular the bracketing, will be explained in Secs. B.2, 3), are confined by all six {1 0 0} monolayers with distances $D_{\{100\}} = 2N\, d_{\{100\}}$ between parallel monolayers (polyhedral NP diameters).

This yields six square shaped {1 0 0} facets with $(N + 1)$ edge atoms each and eight polyhedral atom corners, see Fig. B.1.

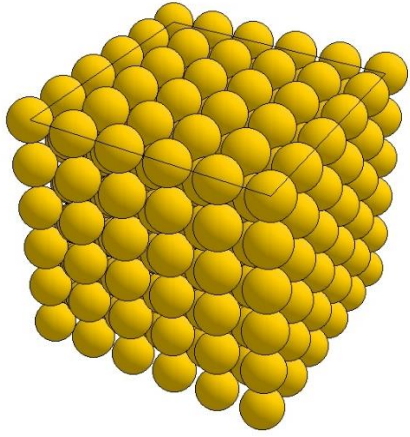

**Figure B.1.** Atom ball model of a generic cubic bcc NP, bcc[5, 0) with an atom at its center. The black lines sketch the square {1 0 0} facet shapes.

(b) **Generic rhombic** bcc NPs, denoted **bcc(0, M]**, are confined by all twelve {1 1 0} monolayers with distances $D_{\{110\}} = 2M\, d_{\{110\}}$ between parallel monolayers (polyhedral NP diameters). This yields rhombic {1 1 0} facets of all twelve monolayers, see Fig. B.2. The {1 1 0} rhombi contain $(M + 1) \times (M + 1)$ atoms each, also described by $2M + 1$ parallel nearest neighbor atom rows of increasing/decreasing length in the sequence 0 (atom), $1a$, …, $M\,a$, …, $1a$, 0 (atom). Altogether, the generic rhombic bcc NPs are described as rhombic dodecahedra offering 14 polyhedral atom corners and reminding of the shape of Wigner-Seitz cells of the face centered cubic (fcc) crystal lattice [a12].

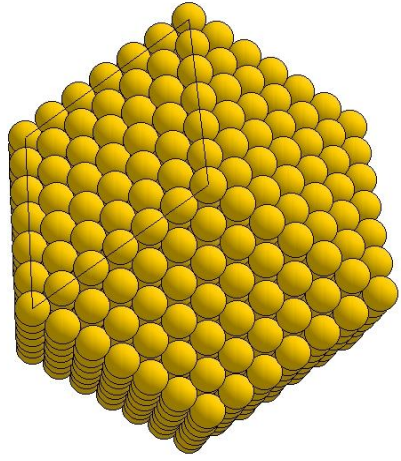

**Figure B.2.** Atom ball model of a generic cubic bcc(0, 6] NP with an atom at its center. The black lines sketch the rhombic {1 1 0} facet shapes.

## B.2. Non-generic bcc Nanoparticles

Non-generic bcc nanoparticles of $O_h$ symmetry are confined by facets with orientations of several netplane families $\{h_i\ k_i\ l_i\}$. This can be considered as combining the confinements of the corresponding generic NPs, $NP_i$, determined by $\{h_i\ k_i\ l_i\}$ with suitable polyhedral diameters, sharing their symmetry center and type (atom or void). As an example we discuss non-generic bcc nanoparticles which combine two generic nanoparticles $NP_1$, $NP_2$ of the types detailed in Sec. B.1. Thus, the resulting bcc NP contains only atoms inside both partner NPs. Here three possible scenarios can be distinguished, nanoparticle $NP_2$ contains all atoms of $NP_1$ (yielding generic $NP_1$), $NP_1$ contains all atoms of $NP_2$ (yielding generic $NP_2$), or $NP_1$ and $NP_2$ share only parts of their atoms (intersecting, which yields true non-generic NPs). These scenarios are determined by the choice of the two netplane families $\{h\ k\ l\}$ and by relationships between the corresponding polyhedral NP diameters $D_{\{hkl\}}$ defined by $N$ and $M$. In the following, we restrict ourselves to non-generic sc NPs defined by pairs of generic cubic and rhombic NPs.

Non-generic bcc NPs contain always an **atom** at their symmetry center. They combine a generic cubic NP, sc[$N$, 0), and a rhombic NP, sc(0, $M$] where polyhedral diameters amount to

$$D_{\{100\}} = 2N\, d_{\{100\}} = N\, a, \qquad D_{\{110\}} = 2M\, d_{\{110\}} = 2M\, a/\sqrt{2} \qquad \text{(B.3)}$$

Thus, the smallest rhombic NP which surrounds the cubic NP is defined with (B.2) by

$$D_{\{110\}} = \sqrt{2}\, D_{\{100\}} \quad \text{and hence} \quad M = N \qquad \text{(B.4)}$$

On the other hand, the smallest cubic NP which surrounds the rhombic NP is defined with (B.2) by

$$D_{\{100\}} = \sqrt{2}\, D_{\{110\}} \quad \text{and hence} \quad N = 2M \tag{B.5}$$

As a consequence, non-generic bcc NPs combining generic cubic and rhombic NPs, $NP_1$, $NP_2$, yield the choices shown in Table B.1.

| $NP_1 / NP_2$ | $NP_1$ in $NP_2$ | $NP_2$ in $NP_1$ |
|---|---|---|
| Cubic / Rhombic bcc[N, 0) / bcc(0, M] | $M \geq N$ | $N \geq 2M$ |

**Table B.1.** Constraints of structure parameters $N$, $M$ for pairs of bcc NPs, $NP_1$ and $NP_2$, sharing an atom as their symmetry center.

Table B.1 shows in particular that true non-generic bcc NPs, defined by pairs of generic cubic and rhombic NPs, bcc[N, 0) and bcc(0, M] refer to structure parameters $N$, $M$ which bracket each other according to

$$M < N < 2M, \quad N/2 < M < N \tag{B.6}$$

These non-generic bcc NPs may be called **cubo-rhombic** and denoted as **bcc[N, M]** combining a generic cubic bcc[N, 0) (first index) with a rhombic NP sc(0, M] (second index). The nomenclature together with the results of Table B.1 suggests an alternative notation for generic NPs where we may formally write

generic cubic     bcc[N, 0) = bcc[N, M] ,     $N \leq M$     (B.7a)

generic rhombic   bcc(0, M] = bcc[N, M] ,     $N \geq 2M$    (B.7b)

## B.3. Cubo-rhombic bcc Nanoparticles

In the following we will discuss the detailed structure of true cubo-rhombic NPs bcc[N, M] determined by corresponding structure parameters $N$, $M$ and confined by the two densest monolayer families {1 0 0} and {1 1 0}. These NPs contain always an **atom** at there symmetry center.

Starting from a generic cubic NP bcc[N, 0) with a polyhedral NP diameter $D_{\{100\}} = 2N\, d_{\{100\}}$ the smallest enclosing rhombic NP bcc(0, M] has a polyhedral NP diameter $D_{\{110\}} = 2M\, d_{\{110\}}$ with $M = N$ according to Table B.1. Shrinking the rhombic bcc NP, i.e. for smaller $M$ values given by $M = N - m$, $m \geq 0$, the cubic and rhombic bcc NPs intersect yielding a **cubo-rhombic** NP bcc[N, M = N - m]. In the following we define this NP by **sc[N, m)** which is compatible with the initial definition of a generic cubic bcc NP denoted sc[$N$, 0) in Sec. B.1. The constraints on $M$ given in Table B.1 can be expressed by equivalent constraints on $m$ yielding

$$N/2 \leq M \leq N \;, \qquad 0 \leq m \leq N/2 \qquad (B.8)$$

where $m$ values larger than $N/2$ result in generic rhombic bcc NPs. The building scheme corresponds to the generic cubic NP bcc[$N$, 0) being truncated at its twelve edges by removing $m$ {1 1 0} facet layers each.

On the other hand, starting from a generic rhombic NP, bcc(0, $M$], with a polyhedral NP diameter $D_{\{110\}} = 2M\, d_{\{110\}}$ the smallest enclosing cubic bcc NP has a polyhedral NP diameter $D_{\{100\}} = 2N\, d_{\{100\}}$ with $N = 2M$ according to Table B.1. Shrinking the cubic bcc NP, i.e. for smaller $N$ values given by $N = 2M - n$ with $n \geq 0$, the cubic and rhombic NPs intersect yielding also a **cubo-rhombic** NP bcc[$N = 2M - n$, $M$]. In the following we define this NP by **bcc($n$, $M$]** which is compatible with the initial definition of a generic rhombic bcc NP denoted bcc(0, $M$] in Sec. A.1. The constraints on $N$ given in Table B.1 can be expressed by equivalent constraints on $n$ yielding

$$M \leq N \leq 2M \;, \qquad 0 \leq n \leq M \qquad \text{cubo-rhombic A } (N \text{ even}) \qquad (B.9)$$

where $n$ values larger than $M$ result in generic cubic bcc NPs. The building scheme corresponds to the generic rhombic NP, sc(0, $M$], being truncated at its six 4-fold symmetry corners by removing $n$ {1 0 0} facet layers each.

The two building schemes must yield the same cubo-rhombic NP bcc[$N$, $M$] where the two structure parameters $N$, $M$ quantify the polyhedral NP diameters $D_{\{100\}}$ and $D_{\{110\}}$ according to

$$D_{\{100\}} = 2N\, d_{\{100\}} \;, \qquad D_{\{110\}} = 2M\, d_{\{110\}} \qquad (B.10)$$

with the basic netplane distances $d_{\{100\}}$ and $d_{\{110\}}$ given by (B.2). Further, the relations

$$M = N - m \;, \qquad N = 2M - n \qquad (B.11a)$$

from the two building schemes can be converted to

$$n = 2M - N \;, \qquad m = N - M \qquad (B.11b)$$
$$N = n + 2m \;, \qquad M = n + m \qquad (B.11c)$$

This suggest an alternative set of structure parameters $n$, $m$ which characterize the deviation of the cubo-rhombic NP bcc[$N$, $M$] from its generic envelope NPs, cubic bcc[$N$, 0) and rhombic bcc(0, $M$]. Thus, **cubo-rhombic** bcc NPs can also be denoted **bcc($n$, $m$)**.

The structure parameters *n*, *m* are also connected with geometric properties of the NP facets, in particular, with characteristic lengths of the facet edges. The cubo-rhombic bcc NPs, bcc[*N*, *M*] ≡ bcc(*n*, *m*), are confined by all six {1 0 0} facets of square shape with $(n + 1) \times (n + 1)$ atoms each and by all twelve {1 1 0} facets, see Fig. B.3. The {1 1 0} facets are hexagonal and contain $2m + 1$ parallel nearest neighbor atom rows of increasing/decreasing length in the sequence $n\,a$, $(n+1)a$, …, $(n + m)a$, …, $(n+1)a$, $n\,a$. Altogether, the bcc NPs are described by cubo-rhombic polyhedra offering 32 ($n > 0$, $m > 0$) corners with true cubic ($n > 0$, $m = 0$, eight corners) and true rhombic ($n = 0$, $m > 0$, 14 corners) being special cases.

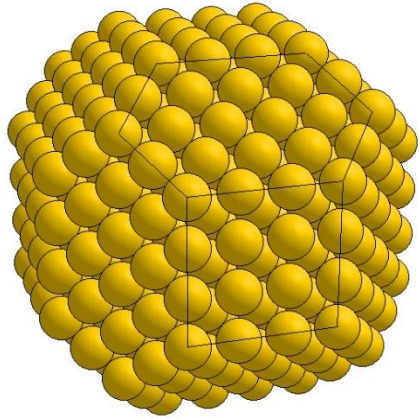

**Figure B.3.** Atom ball model of the cubo-rhombic sc[7, 5] ≡ bcc(3, 2) NP. The black lines sketch the square {1 0 0} and hexagonal {1 1 0} facet shapes.

The confinement of a cubo-rhombic bcc[*N*, *M*] ≡ bcc(*n*, *m*) NP by its generic envelope bcc NPs, cubic bcc[*N*, 0) and rhombic bcc(0, *M*] with *N*, *M* defined by B.11, is illustrated in Fig. B.4 for bcc[10, 7] ≡ bcc(4, 3) with bcc[10, 0] and bcc(0, 7] forming the envelope bcc NPs.

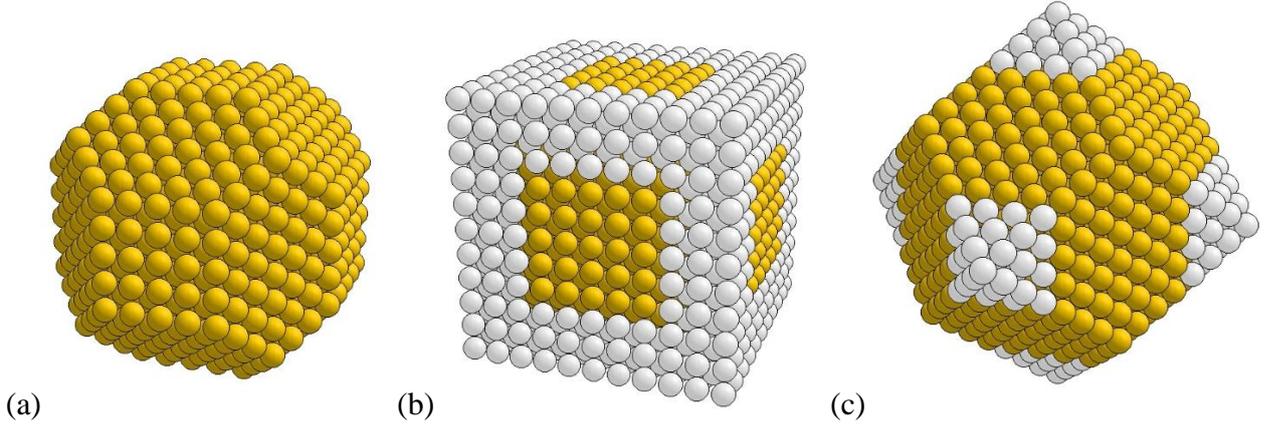

**Figure B.4.** Atom ball models of a cubo-rhombic bcc[10, 7] ≡ bcc(4, 3) NP with its confining generic cubic and rhombic NPs. (a) Initial bcc(4, 3) shown by yellow balls, (b) bcc(4, 3) inside generic bcc[10, 0), and (c) bcc(4, 3) inside generic bcc(0, 7]. The atoms outside bcc(4, 3) are shown by white balls.

Altogether, the equivalence of structure parameters $n$, $m$ and $N$, $M$ according to (B.11) suggests alternative notations of cubo-rhombic bcc NPs introduced earlier. Apart from a notation by bcc[$N$, $M$], meaningful alternatives are bcc($n$, $m$), bcc[$N$, $m$), and bcc($n$, $M$] where the different bracketing denotes the different meanings of the two parameters in the notation. Table B.2 lists types and notations of all intersecting cubo-rhombic bcc NPs.

| Type | bcc($n$, $m$) | bcc[$N$, $M$] | bcc[$N$, $m$) | bcc($n$, $M$] |
|---|---|---|---|---|
| Cubo-rhombic | $n > 0, m > 0$ | $M < N < 2M$ | $0 < m < N/2$ | $0 < n < M$ |
| Generic cubic | $n > 0, m = 0$ | $N = M$ | $m = 0$ | $n = M$ |
| Generic rhombic | $n = 0, m > 0$ | $N = 2M$ | $m = N/2$ | $n = 0$ |
| Atom (A) | $n = m = 0$ | $N = M = 0$ | $N = m = 0$ | $n = M = 0$ |

**Table B.2.** Types and notations of all intersecting and generic bcc NPs.

## B.4. Properties of Cubo-rhombic bcc Nanoparticles

Structural properties characterizing cubo-rhombic bcc nanoparticles can be obtained by geometric consideration and simple algebra where we quote only results of the most important quantities. Structure parameters $n$, $m$ and $N$, $M$, according to (B.11), i.e.

$$n = 2M - N, \qquad m = N - M \qquad \text{(B.11b)}$$
$$N = n + 2m, \qquad M = n + m \qquad \text{(B.11c)}$$

with constraints listed in Table B.1 will be used interchangeably and in mixed combinations to simplify formal expressions. Note that according to (B.11) structure parameters $n$ and $N$ will always be both even or both odd numbers.

The number of atoms forming a bcc($n$, $m$) NP are given by the **volume count** denoted $N_{vol}(n, m)$ where simple algebra yields

$$\begin{aligned} N_{vol}(n, m) &= \frac{1}{2}(2n + 2m + 1)\left[(2n + 2m + 1)^2 + 1\right] - n(n+1)(2n+1) \\ &= \frac{1}{2}(2M + 1)\left[(2M + 1)^2 + 1\right] - n(n+1)(2n+1) \end{aligned} \qquad \text{(B.12)}$$

with **special cases**

Generic cubic
$$\begin{aligned} N_{vol}(n, 0) &= (2n+1)\left[n(n+1)+1\right] \\ &= (4M - 2N + 1)\left[(2M - N)(2M - N + 1) + 1\right] \end{aligned} \qquad \text{(B.13)}$$

Generic rhombic
$$\begin{aligned} N_{vol}(0, m) &= (2m+1)\left[2m(m+1)+1\right] \\ &= (2(N - M) + 1)\left[2(N - M)(N - M + 1) + 1\right] \end{aligned} \qquad \text{(B.14)}$$

The number of atoms on the outermost facets of a bcc($n$, $m$) NP (forming the outer polyhedral shell confining the NP) are given by the **facet count** denoted $N_{shell}(n, m)$ where simple algebra yields

$$N_{shell}(n,m) = 12(n+m)^2 - 6n^2 + 2 = 12M^2 - 6(2M - N)^2 + 2 \qquad \text{(B.15)}$$

with **special cases**

Generic cubic
$$\begin{aligned} N_{shell}(n, 0) &= 6n^2 + 2 \\ &= 6(2M - N)^2 + 2 \qquad n > 0 \end{aligned} \qquad \text{(B.16)}$$

Generic rhombic
$$\begin{aligned} N_{shell}(0, m) &= 12m^2 + 2 \\ &= 12(N - M)^2 + 2 \qquad m > 0 \end{aligned} \qquad \text{(B.17)}$$

The distances between the center of a bcc(*n*, *m*) NP and its outer facet midpoints (**facet distances**) are given by $R_{\{hkl\}}(n, m)$, $\{h\,k\,l\} = \{1\,0\,0\}, \{1\,1\,0\}$, and $\{1\,1\,1\}$, which are connected with the polyhedral NP diameters $D_{\{hkl\}}$ according to

$$R_{\{hkl\}}(n,m) = \frac{1}{2} D_{\{hkl\}} \tag{B.18}$$

Thus, with (B.2), (B.3), (B.11) and *a* denoting the lattice constant of the bcc lattice we obtain

$$R_{\{100\}}(n,m) = \frac{1}{2}(2N)d_{\{100\}} = \frac{a}{2}(n+2m) \tag{B.19}$$

$$R_{\{110\}}(n,m) = \frac{1}{2}(2M)d_{\{110\}} = \frac{a}{\sqrt{2}}(n+m) \tag{B.20}$$

$$R_{\{111\}}(n,m) = \frac{1}{2}(6M)d_{\{111\}} = \frac{\sqrt{3}}{2}a(n+m) \tag{B.21}$$

   +  Note that all $\{1\,1\,1\}$ facets reduce to one atom. Thus, distances $R_{\{111\}}(n, m)$ refer to corner rather than facet distances.

The distances between the center of a bcc(*n*, *m*) NP and its two different types of corners (**corner distances**) are given by $R_{c1}(n, m)$ (referring to each of six ($n = 0$) or 24 ($n > 0$) corners at $\{1\,0\,0\}$ facets) and by $R_{c2}(n, m)$ (referring to each of eight corners at $\{1\,1\,0\}$) where, with (B.11) and *a* denoting the lattice constant of the bcc lattice

$$R_{c1}(n,m) = \frac{a}{2}\sqrt{(n+2m)^2 + 2n^2} = \frac{a}{2}\sqrt{N^2 + 2(2M-N)^2} \tag{B.22}$$

$$R_{c2}(n,m) = \frac{\sqrt{3}}{2}a(n+m) = \frac{\sqrt{3}}{2}aM = R_{\{111\}}(n,m) \tag{B.23}$$

The areas of each of the $\{1\,0\,0\}$ and $\{1\,1\,0\}$ facets of a bcc(*n*, *m*) NP, measured by corresponding corner atoms (**facet areas**), are given by $A_{\{100\}}(n, m)$ (referring to each of six facets) and $A_{\{110\}}(n, m)$ (referring to each of twelve facets) where, with *a* denoting the lattice constant of the bcc lattice

$$A_{\{100\}}(n,m) = a^2 n^2 = a^2 (2M-N)^2 \tag{B.24}$$

$$A_{\{110\}}(n,m) = \frac{1}{\sqrt{2}}a^2 m(2n+m) = \frac{1}{\sqrt{2}}a^2 (N-M)(3M-N) \tag{B.25}$$

## C. Nanoparticles with Face Centered Cubic (fcc) Lattice Structure

The face centered cubic (fcc) lattice can be defined as a non-primitive simple cubic lattice by lattice vectors $\underline{R}_1$, $\underline{R}_2$, $\underline{R}_3$ in Cartesian coordinates together with four lattice basis vectors $\underline{r}_1$ to $\underline{r}_4$ according to

$$\underline{R}_1 = a(1,0,0) \ , \quad \underline{R}_2 = a(0,1,0) \ , \quad \underline{R}_3 = a(0,1,0) \tag{C.1a}$$

$$\underline{r}_1 = a(0,0,0) \ , \quad \underline{r}_2 = a/2\,(0,1,1) \ , \quad \underline{r}_3 = a/2\,(1,0,1) \ , \quad \underline{r}_4 = a/2\,(1,1,0) \tag{C.1b}$$

where $a$ is the lattice constant. The two densest monolayer families of the fcc lattice are described by six square shaped {1 0 0} and eight hexagonal {1 1 1} (highest density) netplanes where distances between adjacent parallel netplanes are given by

$$d_{\{100\}} = a/2 \ , \quad d_{\{111\}} = a/\sqrt{3} \ , \quad d_{\{110\}} = a/(2\sqrt{2}) \tag{C.2}$$

The point symmetry of the fcc lattice is characterized by $O_h$ with symmetry centers at all atom sites and at the void centers of each elementary cell.

Compact face centered cubic nanoparticles (NPs) are confined by finite sections of monolayers (facets) whose structure is described by different netplanes ($h\,k\,l$). If they exhibit central $O_h$ symmetry and show an ($h\,k\,l$) oriented facet they must must also include all other symmetry related facets characterized by orientations of the complete {$h\,k\,l$} family. Thus, general fcc NPs of $O_h$ symmetry are determined by facets whose orientation can be defined by those of different {$h\,k\,l$} families. (As an example, we mention the {1 1 1} family with its eight netplane orientations (±1 ±1 ±1).) Further, according to the symmetry of the fcc host lattice possible NP centers can only be atom or $O_h$ symmetry void sites of the lattice. This will be discussed at different levels of complication in the following.

### C.1. Generic fcc Nanoparticles

First, we consider generic fcc nanoparticles (NPs) of $O_h$ symmetry which are confined by facets with orientations of only one netplane family {$h\,k\,l$}. This allows to distinguish between different generic NP types where we focus on those confined by {1 0 0} or {1 1 1} facets which correspond to the densest monolayers of the fcc lattice and offer the flattest NP facets.

(a) **Generic cubic** fcc NPs are confined by all six {1 0 0} monolayers with distances $D_{\{100\}} = 2N\, d_{\{100\}}$ between parallel monolayers (polyhedral NP diameters). Here we distinguish two NP types.

**Generic cubic A** fcc NPs, denoted **fcc[N, 0)** (the notation of fcc NPs, in particular the bracketing, will be explained in Secs. C.2, 3) contain an **atom** at their symmetry center for **even** $N$ and a **void** for **odd** $N$. They are confined by true square {1 0 0} facets. This yields facets with $(N + 1)$ edge atoms each and eight polyhedral atom corners, see Fig. C.1a.

**Generic cubic B** fcc NPs, denoted **fcc[N, 1)** contain an **atom** at their symmetry center for **odd** $N$ and a **void** for **even** $N$. They are confined by capped square (octagonal) {1 0 0} facets with long edges of N atoms each and eight polyhedral atom corners capped by {1 1 1} microfacets, see Fig. C.1b. .

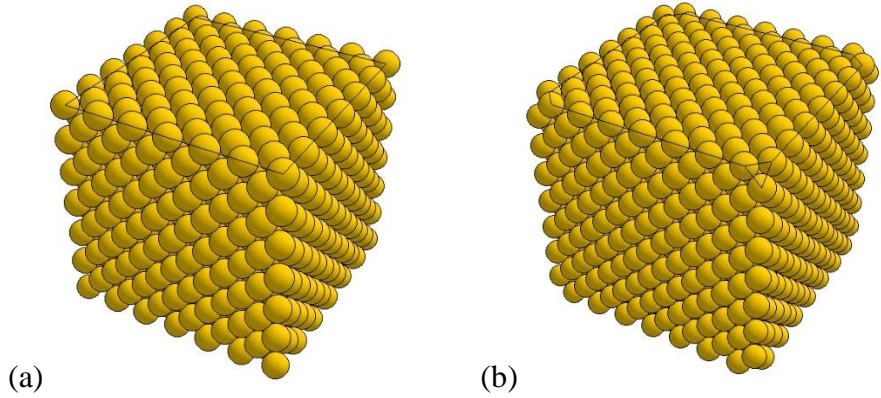

(a)  (b)

**Figure C.1.** Atom ball models of generic cubic fcc NPs, (a) cubic A fcc[6, 0) and (b) cubic B fcc[7, 1). The black lines sketch the square and octagonal {1 0 0} as well as the triangular {1 1 1} facet shapes.

(b) **Generic octahedral** fcc NPs are confined by all eight {1 1 1} monolayers with distances $D_{\{111\}} = K\, d_{\{111\}}$ between parallel monolayers (polyhedral NP diameters). These NPs contain an **atom** at their symmetry center for **even** $K$ and a void **void** for **odd** $K$. Here we distinguish two NP types.

**Generic octahedral A** fcc NPs, denoted **fcc(0, K]**, are confined by true triagonal facets with $(K + 1)$ edge atoms each six polyhedral atom corners, see Fig. C.2a.

**Generic octahedral B** fcc NPs, denoted **fcc(1, K]**, are confined by capped triagonal (hexagonal) {1 1 1} facets with long edges of N - 1 atoms each and six capped polyhedral atom corners with {1 0 0} microfacets. These NPs are not strictly generic since they are derived from generic octahedral A, fcc(0, K] , by removing all six corner atoms. However, they will be useful in the discussion of Sec. C.2 and later.

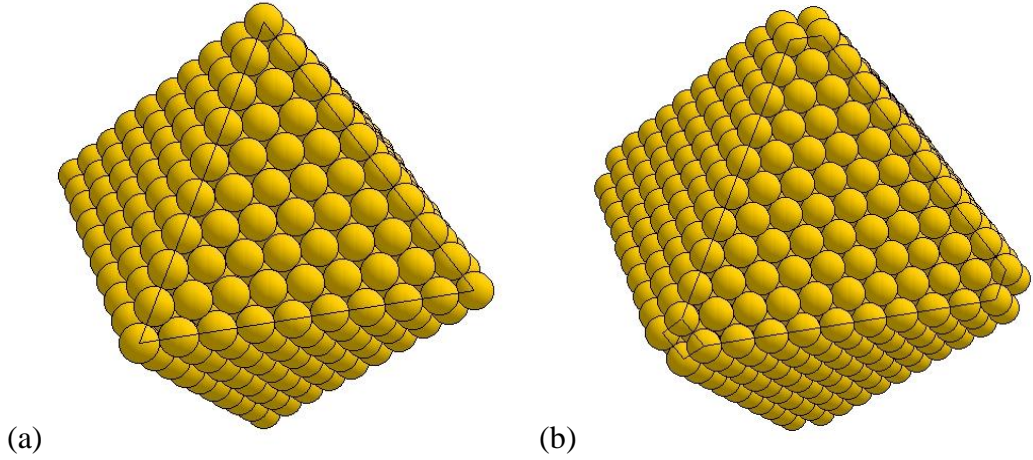

(a)  (b)

**Figure C.2.** Atom ball models of generic octahedral fcc NPs, (a) octahedral A fcc(0, 9] and (b) octahedral B fcc(1, 11] . The black lines sketch the triangular {1 1 1} and square {1 0 0} facet shapes.

**(c)** **Generic cuboctahedral** fcc NPs, denoted **fcc(N, 2N]** (the notation, in particular the bracketing, will be explained in Secs. C.2, 3), contain always an **atom** at their symmetry center. They are confined by the six {1 0 0} monolayers with distances $D_{\{100\}} = 2N\, d_{\{100\}}$ as well as by the eight {1 1 1} monolayers with distances $D_{\{111\}} = 2N\, d_{\{111\}}$ between parallel monolayers (polyhedral NP diameters). They are described by six square shaped {1 0 0} facets with $(N + 1) \times (N + 1)$ atoms each and by eight triangular shaped {1 1 1} facets with $(n + 1)$ atom edges, see Fig. C.3b.

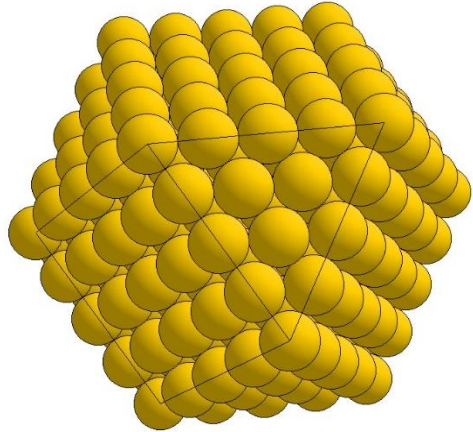

**Figure C.3.** Atom ball model of a generic cuboctahedral NP fcc(4, 8]. The black lines sketch the square {1 0 0} and triangular {1 1 1} facet shapes.

## C.2. Non-generic fcc Nanoparticles

Non-generic fcc nanoparticles of $O_h$ symmetry are confined by facets with orientations of several netplane families $\{h_i\ k_i\ l_i\}$. This can be considered as combining the confinements of the corresponding generic NPs, $NP_i$, determined by $\{h_i\ k_i\ l_i\}$ with suitable polyhedral diameters, sharing their symmetry center and type (atom or void). As an example we discuss non-generic fcc nanoparticles which combine two generic nanoparticles $NP_1$, $NP_2$ of the types detailed in Sec. C.1. Thus, the resulting fcc NP contains only atoms inside both partner NPs. Here three possible scenarios can be distinguished, nanoparticle $NP_2$ contains all atoms of $NP_1$ (yielding generic $NP_1$), $NP_1$ contains all atoms of $NP_2$ (yielding generic $NP_2$), or $NP_1$ and $NP_2$ share only parts of their atoms (intersecting which yields true non-generic NPs). These scenarios are determined by the choice of the two netplane families $\{h\ k\ l\}$ and by relationships between the corresponding polyhedral NP diameters $D_{\{hkl\}}$ defined by $N$ and $K$. In the following, we restrict ourselves to non-generic fcc NPs defined by pairs of generic cubic and octahedral NPs.

Non-generic fcc NPs with an **atom** at their symmetry center combine a generic cubic NP (cubic A, fcc[$N$, 0) with even $N$, or cubic B, fcc[$N$, 1) with odd $N$), with a generic octahedral NP (octahedral A, B, fcc(0, $K$], fcc(1, $K$] with even $K$ ) where polyhedral diameters amount to

$$D_{\{100\}} = 2N\, d_{\{100\}} = N\,a, \quad D_{\{111\}} = K\, d_{\{111\}} = K\,a/\sqrt{3} \tag{C.3}$$

Thus, the smallest octahedral NP which surrounds a cubic A or B NP is defined with (C.2) by

cubic A: $\quad D_{\{111\}} = \sqrt{3}\, D_{\{100\}}$ $\quad$ and hence $\quad K = 3N$ $\quad$ (C.4a)

cubic B: $\quad D_{\{111\}} = \sqrt{3}\, D_{\{100\}} - 2d_{\{111\}}/2$ $\quad$ and hence $\quad K = 3N - 1$ $\quad$ (C.4b)

On the other hand, the smallest cubic NP surrounding an octahedral A, B NP is defined with (C.2) by

octahedral A: $\quad D_{\{100\}} = \sqrt{3}\, D_{\{111\}}$ $\quad$ and hence $\quad N = K$ $\quad$ (C.5a)

octahedral B: $\quad D_{\{100\}} = \sqrt{3}\, D_{\{111\}} - 2d_{\{100\}}$ $\quad$ and hence $\quad N = K - 1$ $\quad$ (C.5b)

As a consequence, non-generic fcc NPs with an atom at their symmetry center combining generic cubic and rhombic NPs, $NP_1 / NP_2$, yield the three choices shown in Table C.1.

| NP$_1$ / NP$_2$ | NP$_1$ in NP$_2$ | NP$_2$ in NP$_1$ |
|---|---|---|
| Cubic A / Octahedral A<br>fcc[*N*, 0) / fcc(0, *K*] | $K \geq 3N$ | $N \geq K$ |
| Cubic B / Octahedral B<br>fcc[*N*, 1) / fcc(1, *K*] | $K \geq 3N - 1$ | $N \geq K - 1$ |
| Cubic A / Octahedral B<br>fcc[*N*, 0) / fcc(1, *K*] | $K \geq 3N$ | $N \geq K$ (<) |
| Cubic B / Octahedral A<br>fcc[*N*, 1) / fcc(0, *K*] | $K \geq 3N - 1$ | $N \geq K + 1$ (<) |

**Table C.1.** Constraints of structure parameters *N*, *M* for pairs of fcc NPs, NP$_1$ and NP$_2$, sharing an atom or void as their symmetry center. Entries labeled (<) refer to combinations NP$_1$ / NP$_2$ where NP$_2$ never touches the surface of NP$_1$.

Non-generic fcc NPs with a **void** at their symmetry center combine a generic cubic NP (cubic A, C, fcc[*N*, 0) with odd *N*, or cubic B, fcc[*N*, 1) with even *N*), with a generic octahedral NP (octahedral A, B, fcc(0, *K*], fcc(1, *K*] with odd *K*) where, in complete analogy to the atom centered NPs, polyhedral diameters are given by (C.3). Further, smallest octahedral NPs surrounding a cubic A or B NPs are defined by (C.4) and smallest cubic NPs surrounding octahedral A, B NPs by (C.5). This shows that Table 1 applies also to NPs with a void at their symmetry center if *N*, *K* are chosed accordingly.

Table C.1 shows in particular that true non-generic fcc NPs, defined by pairs of generic cubic NPs, fcc[*N*, *p*), *p* = 0, 1, and octahedral NPs, fcc(*q*, *K*], *q* = 0, 1, refer to structure parameters *N*, *K* which bracket each other. As examples we mention

$$N < K < 3N \qquad \text{fcc}[N, 0) / \text{fcc}(0, K] \qquad (C.6a)$$
$$N + 1 < K < 3N - 1 \qquad \text{fcc}[N, 1) / \text{fcc}(1, K] \qquad (C.6c)$$

These non-generic fcc NPs may be called **cuboctahedral** and denoted as **fcc[*N*, *M*]** combining a generic cubic (first index) with an octahedral NP (second index) where cubic A NPs combine with octahedral A and cubic B with octahedral B fcc NPs.

The nomenclature together with the results of Table C.1 suggests an alternative notation for generic NPs where we may formally write

$$\text{generic cubic A, B} \qquad \text{fcc}[N, p) = \text{fcc}[N, K] \qquad K \geq 3N - p, \ p = 0, 1, 2 \quad (C.7a)$$
$$\text{generic octahedral A, B} \qquad \text{fcc}(q, K] = \text{fcc}[N, K] \qquad N \geq K - q, \ q = 0, 1 \qquad (C.7b)$$

## C.3. Cuboctahedral fcc Nanoparticles

In the following we will discuss the detailed structure of true cuboctahedral NPs fcc[$N$, $K$] determined by corresponding structure parameters $N$, $K$ and confined by the two densest monolayer families {1 0 0} and {1 1 1}. These NPs contain an **atom** at there symmetry center for **even K** while their symmetry center is **void** for **odd K**.

First, we consider **atom centered** fcc NPs. Starting from a generic cubic A NP fcc[$N$, 0) with even $N$ and a polyhedral NP diameter $D_{\{100\}} = 2N\, d_{\{100\}}$ the smallest enclosing octahedral A NP fcc(0, $K$] has a polyhedral NP diameter $D_{\{111\}} = K\, d_{\{111\}}$ with even $K = 3N$ according to Table C.1. Shrinking the octahedral fcc NP, i.e. for smaller $K$ values given by $K = 3N - k$, $k = 0, 2, 4 \ldots$ the cubic and octahedral fcc NPs intersect yielding a **cuboctahedral** NP fcc[$N$, $K = 3N - k$]. In the following we define this NP by **fcc[$N$, $k$)** which is compatible with the initial definition of a generic cubic fcc NP denoted fcc[$N$, 0) in Sec. C.1. The constraints on $K$ given in Table C.1 can be expressed by equivalent constraints on $k$ yielding

$$N \leq K \leq 3N \, , \qquad 0 \leq k \leq 2N \ \text{with even } k \qquad\qquad \text{(C.8a)}$$

where $k$ values larger than $2N$ result in generic octahedral fcc NPs.

Starting from a generic cubic B NP fcc[$N$, 1) with odd $N$ and a polyhedral NP diameter $D_{\{100\}} = 2N\, d_{\{100\}}$ the smallest enclosing octahedral B NP fcc(1, $K$] has a polyhedral NP diameter $D_{\{111\}} = K\, d_{\{111\}}$ with even $K = 3N - 1$ according to Table C.1. Shrinking the octahedral fcc NP, i.e. for smaller $K$ values given by $K = 3N - k$, $k = 1, 3, 5 \ldots$ the cubic and octahedral fcc NPs intersect yielding also a **cuboctahedral** NP fcc[$N$, $K = 3N - k$] which we define by **fcc[$N$, $k$)**. Here the constraints on $K$ in Table C.1 lead to constraints on $k$ according to

$$N + 1 \leq K \leq 3N - 1 \, , \qquad 1 \leq k \leq 2N - 1 \ \text{with odd } k \qquad\qquad \text{(C.8b)}$$

where $k$ values larger than $2N$ result in generic octahedral fcc NPs.

As a result, the two building schemes for all possible values of $N$ correspond to generic cubic A or B NPs, fcc[$N$, 0) or fcc[$N$, 1), being truncated at their eight edges by removing $k/2$ or $(k - 1)/2$ {1 1 1} facet layers each.

On the other hand, starting from a generic octahedral A NP, fcc(0, $K$] with even $K = 3p$, i.e. being also a multiple of 3 (or with $K = 3p - 2$) and a polyhedral NP diameter $D_{\{111\}} = K\, d_{\{111\}}$ the smallest enclosing cubic A fcc NP fcc[$N$, 0) has a polyhedral NP diameter $D_{\{100\}} = 2N\, d_{\{100\}}$ with even $N = K$ according to Table C.1. Shrinking the cubic fcc NP, i.e. for smaller $N$ values given by $N = K - n$ with $n = 0, 2, 4 \ldots$ the cubic and octahedral NPs intersect yielding also a **cuboctahedral** fcc[$N = K - n, K$] with even $n$. In the following we define this NP by **fcc($n, K$]** which is compatible with the initial definition of a generic octahedral fcc NP denoted sc(0, $K$] in Sec. C.1. The constraints on $N$ given in Table C.1 can be expressed by equivalent constraints on $n$ yielding

$$K/3 \leq N \leq K \ , \qquad 0 \leq n \leq (2/3)K \ , \quad n \text{ even} \qquad (C.9a)$$

where $n$ values larger than $(2/3)K$ result in generic cubic fcc NPs.

For an octahedral B NP fcc(1, $K$] with even $K = 3p - 1$ the smallest enclosing cubic NP is fcc[$N$, 1) and has a polyhedral NP diameter $D_{\{100\}} = 2N\, d_{\{100\}}$ with odd $N = K - 1$ according to Table C.1. Shrinking the cubic fcc NP, i.e. for $N = K - n$ with $n = 1, 3, 5 \ldots$ the cubic and octahedral NPs intersect yielding also a **cuboctahedral** fcc[$N = K - n, K$] defined as **fcc($n, K$]**, however, with odd $n$. This leads to constraints on $n$, together with Table C.1

$$(K + 1)/3 \leq N \leq K - 1 \ , \qquad 1 \leq n \leq (2K - 1)/3 \ , \quad n \text{ odd} \qquad (C.9b)$$

where $n$ values larger than $(2K - 1)/3$ result in generic cubic fcc NPs.

As a result, the building schemes for all possible values of $K$ correspond to generic octahedral A or B NPs, fcc(0, $K$] or fcc(1, $K$], being truncated at their six 4-fold symmetry corners by removing $n$ or $(n - 1)\{1\,0\,0\}$ facet layers each.

The two building scenarios, starting from a generic octahedral or generic cubic NPs, must yield the same cuboctahedral NP fcc[$N, M$] where the two structure parameters $N, K$ quantify the polyhedral NP diameters $D_{\{100\}}$ and $D_{\{111\}}$ according to

$$D_{\{100\}} = 2N\, d_{\{100\}} \ , \qquad D_{\{111\}} = K\, d_{\{11\}} \qquad (C.10)$$

with the basic netplane distances $d_{\{100\}}$ and $d_{\{111\}}$ given by (C.2). Further, the relations

$$N = K - n \ , \qquad K = 3N - k \ , \qquad (C.11a)$$

from the two building schemes can be converted to

$$n = K - N \ , \qquad k = 3N - K \qquad (C.11b)$$
$$2N = n + k \ , \qquad 2K = 3n + k \qquad (C.11c)$$

Next, we consider **void centered** fcc NPs whose treatment is completely analogous to the atom centered case. In fact, all derivations and relationships (C.8a) to (C.11c) are identical with those obtained for atom centered NPs except that all $N$ or $K$ values which are even with one centering will be odd with the other and viceversa.

Altogether, relation (C.11) suggest an alternative set of structure parameters $n$, $k$ which characterize the deviation of the cuboctahedral NP fcc[$N$, $K$] from its generic envelope NPs, cubic fcc[$N$, 0) or fcc[$N$, 1) and rhombic fcc(0, $K$] or fcc(1, $K$]. Thus, **cuboctahedral** NPs can also be denoted **fcc($n, k$)**.

The structure parameters $n$, $k$ are also connected with geometric properties of the NP facets, in particular, with characteristic lengths of the facet edges where we can distinguish two types of cuboctahedral fcc NPs, **truncated octahedral** and **truncated cubic** species.

**Truncated octahedral** NPs, fcc[$N$, $K$] / fcc($n$, $k$) with (C.11) are defined by

$$N \leq K \leq 2N , \quad 0 \leq n \leq N , \quad N \leq k \leq 2N , \quad \text{i.e. } n \leq k \qquad \text{(C.12a)}$$

where the symmetry center is an atom site for even $K$ ($N$, $n$, $k$ all even or all odd) or void for odd $K$ ($N$ even and $n$, $k$ odd or $N$ odd and $n$, $k$ even). These NPs are confined by all six {1 0 0} facets of square shape with $(n + 1) \times (n + 1)$ atoms each and by all eight {1 1 1} facets of hexagonal shape with alternating $(n + 1)$ and $(k - n + 2)/2$ atom edges, see Fig. C.4. The polyhedral NPs offer 24 ($n > 0$, $k > N$) corners with true octahedral ($n = 0$, $k = 2N$, six corners) and generic cuboctahedral ($n = k = N$, twelve corners) being special cases.

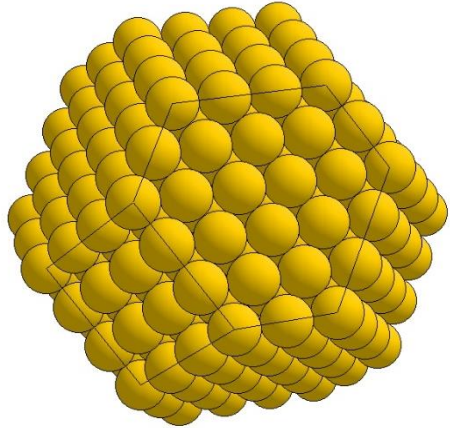

**Figure C.4.** Atom ball model of a truncated octahedral NP fcc[5, 8] ≡ fcc(3, 7) ≡ fcc(3, 2)$_o$. The black lines sketch the square {1 0 0} and hexagonal {1 1 1} facet shapes.

As a result of the building procedure, truncated octahedral fcc[$N$, $K$] NPs are each confined by two generic NPs, one octahedral fcc(0, $K$] or fcc(1, $K$] NP and one cubic fcc($N$, 0) or fcc($N$, 1). This is illustrated for the NP fcc[5, 8] ≡ fcc(3, 2)$_o$ in Fig. C.5.

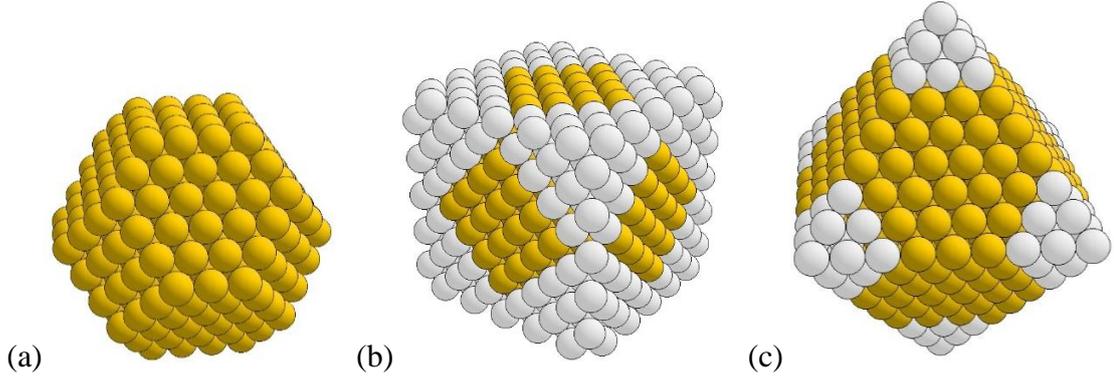

(a) (b) (c)

**Figure C.5.** Atom ball models of a truncated octahedral NP fcc[5, 8] ≡ fcc(3, 2)$_o$ with its confining generic cubic and octahedral NPs. (a) Initial fcc(3, 2)$_o$ shown by yellow balls, (b) fcc(3, 2)$_o$ inside generic fcc[5, 1), and (c) fcc(3, 2)$_o$ inside generic fcc(0, 8]. The atoms outside fcc(3, 2)$_o$ are shown by white balls.

Further, Fig. C.6 shows the generic cuboctahedral NP fcc(4, 8] ≡ fcc(4, 0)$_o$ with its confining generic cubic fcc[4, 0) and octahedral fcc(0, 8] NPs.

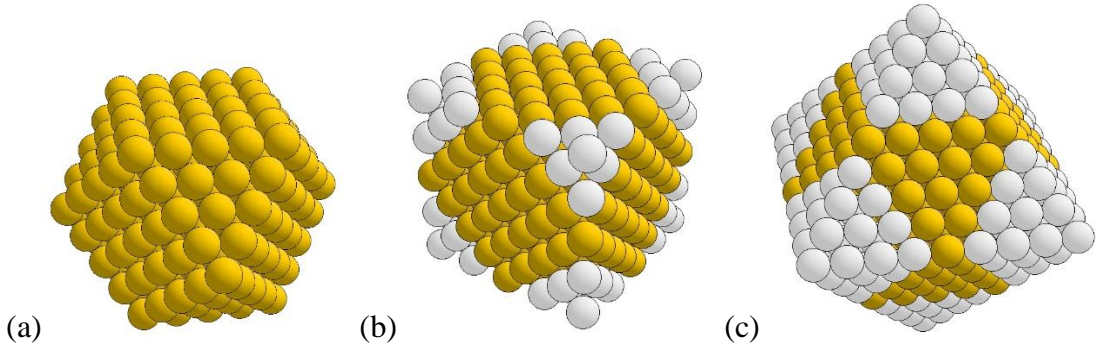

(a) (b) (c)

**Figure C.6.** Atom ball models of a cuboctahedral NP fcc[4, 8] ≡ fcc(4, 0)$_o$ ≡ fcc(0, 4)$_c$ with its confining generic cubic and octahedral NPs. (a) Initial fcc(4, 0)$_o$ shown by yellow balls, (b) fcc(4, 0)$_o$ inside generic fcc[4, 0), and (c) fcc(4, 0)$_o$ inside generic fcc(0, 8]. The atoms outside fcc(4, 0)$_o$ are shown by white balls.

**Truncated cubic** NPs, fcc[$N$, $K$] / fcc($n$, $k$) with (C.11) are defined by

$$2N \leq K \leq 3N, \quad N \leq n \leq 2N, \quad 0 \leq k \leq N, \quad \text{i.e. } k \leq n \quad \text{(C.12b)}$$

where the symmetry center is an atom site for even $K$ ($N$, $n$, $k$ all even or all odd) or void for odd $K$ ($N$ even and $n$, $k$ odd or $N$ odd and $n$, $k$ even). They are confined by all six {1 0 0} facets of octagonal shape with alternating ($n - k + 2$)/2 and ($k + 1$) atom edges and by all eight {1 1 1} facets of triangular shape with edges of $k + 1$ atoms each, see Fig. C.7. The polyhedral NPs offer 24 ($n > 0$, $k > 0$) corners with cuboctahedral ($n = 0$, $k > 0$, twelve corners) and true cubic ($n > 0$, $k = 0$, eight corners) being special cases.

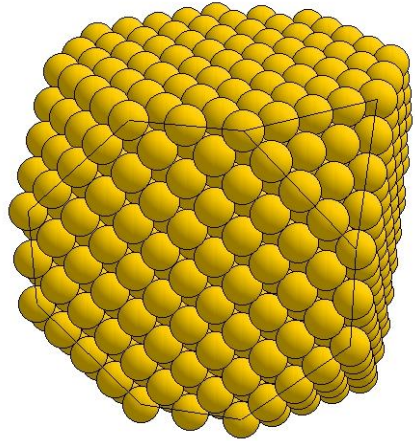

**Figure C.7.** Atom ball model of a truncated cubic NP, fcc[6, 14] ≡ fcc(8, 4) ≡ fcc(2, 4)$_c$. The black lines sketch the octagonal {1 0 0} and triangular {1 1 1} facet shapes.

As a result of the building procedure, truncated cubic fcc[$N$, $K$] NPs are each confined by two generic NPs, one octahedral fcc(0, $K$] or fcc(1, $K$] NP and one cubic fcc($N$, 0) or fcc($N$, 1). This is illustrated for the NP fcc[6, 14] ≡ fcc(2, 4)$_c$ in Fig. C.8.

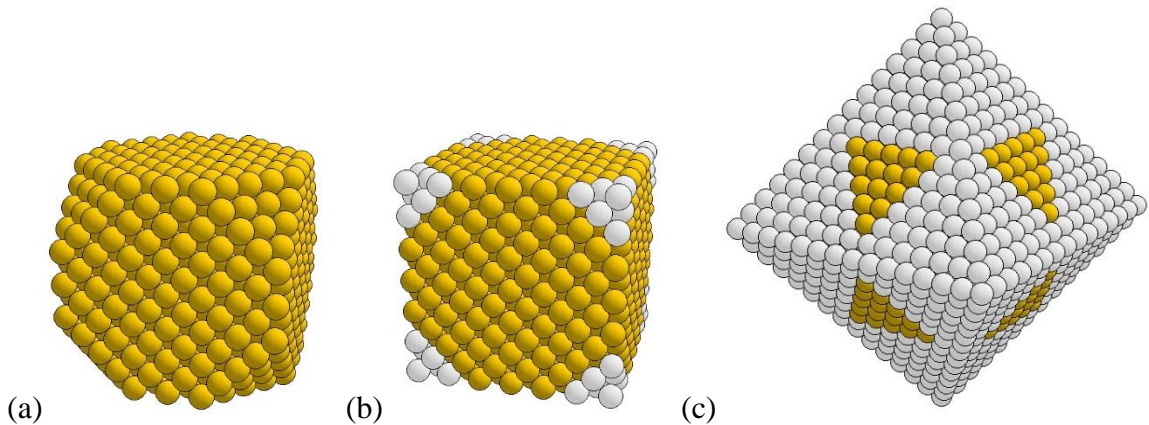

(a)  (b)  (c)

**Figure C.8.** Atom ball models of a cuboctahedral NP fcc[6, 14] ≡ fcc(2, 4)$_c$ with its confining generic cubic and octahedral NPs. (a) Initial fcc(2, 4)$_c$ shown by yellow balls, (b) fcc(2, 4)$_c$ inside generic fcc[6, 0], and (c) fcc(2, 4)$_c$ inside generic fcc(0, 14]. The atoms outside fcc(2, 4)$_c$ are shown by white balls.

Altogether, the equivalence of structure parameters $n$, $k$ and $N$, $K$ according to (C.11) suggests alternative notations of cuboctahedral fcc NPs introduced earlier. Apart from a notation by fcc[$N$, $K$], meaningful alternatives are fcc($n$, $k$), fcc[$N$, $k$), and fcc($n$, $K$] where the different bracketing denotes the different meanings of the two parameters in the notation. Table C.2 lists types and notations of all cuboctahedral fcc NPs where atom centered NPs refer to even $K$ values while for void centered NPs $K$ values are odd.

| Type | fcc[N, M] | fcc(n, k) | fcc[N, k] | fcc(n, K] |
|---|---|---|---|---|
| Cuboctahedral, truncated octahedral | $N \leq K \leq 2N$<br>$n \leq k$ | $0 \leq n \leq N$<br>$N \leq k \leq 2N$ | $N \leq k \leq 2N$ | $0 \leq n \leq N$ |
| Cuboctahedral, truncated cubic | $2N \leq K \leq 3N$<br>$n \geq k$ | $N \leq n \leq 2N$<br>$0 \leq k \leq N$ | $0 \leq k \leq N$ | $N \leq n \leq 2N$ |
| Cuboctahedral, generic | $K = 2N$<br>$n = k$ | $n = k = N$ | $k = N$ | $n = K/2$ |
| Generic cubic A | $K = 3N$ | $n > 0, k = 0$ | $k = 0$ | $n = K/3$ |
| Generic cubic B | $K = 3N - 1$ | $n > 0, k = 1$ | $k = 1$ | $n = (K - 2)/3$ |
| Generic octahedral A | $N = K$ | $n = 0, k > 0$ | $m = N/2$ | $n = 0$ |
| Generic octahedral B | $N = K - 1$ | $n = 1, k > 0$ | $m = (N-1)/2$ | $n = 1$ |
| Elementary cube (B) | $N = 1, K = 3$ | $n = 1, k = 0$ | $N = 1, k = 0$ | $n = 1, K = 3$ |
| Elementary oct. (B) | $N = K = 1$ | $n = 0, k = 1$ | $N = 1, k = 1$ | $n = 0, K = 1$ |
| Atom (A) | $N = K = 0$ | $n = k = 0$ | $N = k = 0$ | $n = K = 0$ |

**Table C.2.** Types and notations of all intersecting and generic fcc NPs.

## C.4. Properties of Cuboctahedral fcc Nanoparticles

Structural properties characterizing cuboctahedral fcc nanoparticles can be obtained by geometric consideration and simple algebra where we quote only results of the most important quantities. Structure parameters $n$, $k$ and $N$, $K$, according to (C.11), i.e.

$$n = K - N , \qquad k = 3N - K$$
$$2N = n + k , \qquad 2K = 3n + k$$

with constraints listed in Tables C.1, 2 will be used interchangeably and in mixed combinations to simplify formal expressions. Note that according to (C.11c) structure parameters $n$ and $k$ will always be both even or both odd numbers.

As discussed in Section C.3, there are two types of cuboctahedral fcc nanoparticles, those of truncated octahedral and truncated cubic type, which are described by different shapes and, therefore different parameter expressions.

**Truncated octahedral** fcc NPs are determined by (C.12a)

$$N \leq K \leq 2N , \quad 0 \leq n \leq N , \quad N \leq k \leq 2N , \quad \text{i.e. } n \leq k$$

For geometric convenience we define an alternative structure parameter

$$k' = k - N \tag{C.13}$$

which transforms relations (C.11) to

$$n = K - N, \qquad k' = 2N - K \tag{C.14a}$$
$$N = n + k', \qquad K = 2n + k' \tag{C.14b}$$

allowing simplifications in some of the subsequent formulas. This will be expressed by an alternative nomenclature of these NPs as fcc($n$, $k$) = **fcc($n$, $k'$)$_o$**. Note that according to (C14b) structure parameters $K$ and $k'$ will always be both even or both odd numbers.

**Truncated cubic** fcc NPs are determined by (C.12b)

$$2N \leq K \leq 3N, \quad N \leq n \leq 2N, \quad 0 \leq k \leq N, \quad \text{i.e. } k \leq n$$

For geometric convenience we define an alternative structure parameter

$$n' = n - N \tag{C.15}$$

which transforms relations (C.11) to

$$n' = K - 2N, \qquad k = 3N - K \tag{C.16a}$$
$$N = n' + k, \qquad K = 3n' + 2k \tag{C.16b}$$

allowing simplifications in some of the subsequent formulas. This will be expressed by an alternative nomenclature of these NPs as fcc($n$, $k$) = **fcc($n'$, $k$)$_c$**. Note that according to (C.16a) structure parameters $K$ and $n'$ will always be both even or both odd numbers.

The two types of fcc NPs will be discussed separately in the following.

## C.4a. Cuboctahedral fcc Nanoparticles, Truncated Octahedral Type

The number of atoms forming a truncated octahedral fcc($n$, $k$) NP are given by the **volume count** denoted $N_{vol}(n, k)$ where simple algebra yields

$$\begin{aligned} N_{vol}(n,k) &= \frac{1}{12}(k+3n+2)\left[(k+3n+2)^2+2\right] - n(n+1)(2n+1) \qquad k \geq n \\ &= \frac{1}{3}(K+1)\left[2(K+1)^2+1\right] - (K-N)(K-N+1)\left[2(K-N)+1\right] \end{aligned} \tag{C.17}$$

with **special cases**

$$\text{Generic octahedral A} \qquad N_{vol}(0,k) = \frac{1}{12}(k+2)\left[(k+2)^2+2\right]$$
$$= \frac{1}{3}(N+1)\left[2(N+1)^2+1\right] \tag{C.18}$$

| | | |
|---|---|---|
| Generic octahedral B | $N_{vol}(1,k) = \frac{1}{12}(k+5)\left[(k+5)^2 + 2\right] - 6$ | |
| | $= \frac{1}{3}(N+2)\left[2(N+2)^2 + 1\right] - 6$ | (C.19) |
| Generic cuboctahedral | $N_{vol}(n,n) = (2n+1)\left[\frac{5}{3}n(n+1) + 1\right]$ | |
| | $= (2N+1)\left[\frac{5}{3}N(N+1) + 1\right]$ | (C.20) |

The number of atoms on the outermost facets of a truncated octahedral fcc(n, m) NP (forming the outer polyhedral shell confining the NP) are given by the **facet count** denoted $N_{shell}(n, m)$ where simple algebra yields

$$N_{shell}(n,k) = (k+3n)^2 - 6n^2 + 2 = 4K^2 - 6(K-N)^2 + 2 \qquad k \geq n \tag{C.21}$$

with **special cases**

| | | |
|---|---|---|
| Generic octahedral A | $N_{shell}(0,k) = k^2 + 2 = 4N^2 + 2$ | (C.22) |
| Generic octahedral B | $N_{shell}(1,k) = (k+3)^2 - 4 = 4(N+1)^2 - 4$ | (C.23) |
| Generic cuboctahedral | $N_{shell}(n,n) = 10n^2 + 2 = 10N^2 + 2$ | (C.24) |

The distances between the center of a truncated octahedral fcc(n, k) NP and its outer facet midpoints (**facet distances**) are given by $R_{\{hkl\}}(n, k)$, $\{h\, k\, l\} = \{1\, 1\, 1\}, \{1\, 0\, 0\}$ and which are connected with the polyhedral NP diameters $D_{\{hkl\}}$ according to

$$R_{\{hkl\}}(n,m) = \frac{1}{2}D_{\{hkl\}} \tag{C.25}$$

Thus, with (C.2), (C.3), (C.11) and $a$ denoting the lattice constant of the fcc lattice we obtain

$$R_{\{111\}}(n,k) = (K/2)d_{\{111\}} = \frac{a}{4\sqrt{3}}(3n+k) \qquad k \geq n \tag{C.26}$$

$$R_{\{100\}}(n,k) = N\,d_{\{100\}} = \frac{a}{4}(n+k) \tag{C.27}$$

The distances between the center of a truncated octahedral fcc(n, k) NP and its corners (**corner distances**) are given by $R_c(n, k)$ (referring to each of six ($k = n$) or 24 ($k > n$) corners at {1 0 0} facets) where with (C.11) and $a$ denoting the lattice constant of the fcc lattice

$$R_c(n,k) = \frac{a}{4}\sqrt{(k+n)^2 + 4n^2} = \frac{a}{2}\sqrt{N^2 + (K-N)^2} \qquad k \geq n \tag{C.28}$$

The areas of each of the {1 1 1} and {1 0 0} facets of a truncated octahedral fcc($n$, $k$) NP, measured by corresponding corner atoms (**facet areas**), are given by $A_{\{111\}}(n, k)$ (referring to each of eight hexagonal (triangular) {1 1 1} facets) and $A_{\{100\}}(n, k)$ (referring to each of six {1 0 0} facets) where, with (C.2), (C.3), (C.11) and $a$ denoting the lattice constant of the fcc lattice

$$A_{(111)}(n, k) = \frac{\sqrt{3}}{8} a^2 \left[ \frac{1}{4}(k + 3n)^2 - 3n^2 \right] = \frac{\sqrt{3}}{8} a^2 \left[ K^2 - 3(K - N)^2 \right] \quad \text{(C.29)}$$

$$A_{(100)}(n, k) = \frac{1}{2} a^2 n^2 = \frac{1}{2} a^2 (K - N)^2 \quad \text{(C.30)}$$

## C.4b. Cuboctahedral fcc Nanoparticles, Truncated Cubic Type

The number of atoms forming a truncated cubic fcc($n$, $k$) NP are given by the **volume count** denoted $N_{vol}(n, k)$ where simple algebra yields

$$\begin{aligned} N_{vol}(n,k) &= \frac{1}{2}(k + n + 2)\left[(k + n)(k + n + 1) + 1\right] - \frac{1}{3}k(k + 2)(2k - 1) \quad n \geq k \\ &= (N + 1)\left[2N(2N + 1) + 1\right] - \frac{1}{3}(3N - K)(3N - K + 2)(2(3N - K) - 1) \end{aligned} \quad \text{(C.31)}$$

with **special cases**

Generic cubic A
$$\begin{aligned} N_{vol}(n,0) &= \frac{1}{2}(n + 2)\left[n(n + 1) + 1\right] \\ &= (N + 1)\left[2N(2N + 1) + 1\right] \end{aligned} \quad \text{(C.32)}$$

Generic cubic B
$$\begin{aligned} N_{vol}(n,1) &= \frac{1}{2}(n + 3)\left[(n + 1)(n + 2) + 1\right] - 1 \\ &= (N + 1)\left[2N(2N + 1) + 1\right] - 1 \end{aligned} \quad \text{(C.33)}$$

Generic cuboctahedral
$$\begin{aligned} N_{vol}(n,n) &= (2n + 1)\left[\frac{5}{3}n(n + 1) + 1\right] \\ &= (2N + 1)\left[\frac{5}{3}N(N + 1) + 1\right] \end{aligned} \quad \text{(C.34)}$$

The number of atoms on the outermost facets of a truncated cubic fcc($n$, $m$) NP (forming the outer polyhedral shell confining the NP) are given by the **facet count** denoted $N_{shell}(n, m)$ where simple algebra yields

$$N_{shell}(n, k) = 3(k + n)^2 - 2(k^2 - 1) = 12N^2 - 2\left[(3N - K)^2 - 1\right] \quad n \geq k \quad \text{(C.35)}$$

with **special cases**

Generic cubic A
$$N_{shell}(n, 0) = 3n^2 + 2 = 3(K - N)^2 + 2 \quad \text{(C.36)}$$

Generic cubic B $\qquad N_{shell}(n,1) = 3(n+1)^2 = 12N^2 \qquad$ (C.37)

Generic cuboctahedral $\qquad N_{shell}(n,n) = 10n^2 + 2 = 10N^2 + 2 \qquad$ (C.38)

The distances between the center of a truncated cubic fcc(n, k) NP and its outer facet midpoints (**facet distances**) are given by $R_{\{hkl\}}(n, k)$, $\{h\,k\,l\} = \{1\,1\,1\}, \{1\,0\,0\}$ and which are connected with the polyhedral NP diameters $D_{\{hkl\}}$ according to

$$R_{\{hkl\}}(n,m) = \frac{1}{2} D_{\{hkl\}} \qquad (C.39)$$

Thus, with (C.2), (C.3), (C.11) and $a$ denoting the lattice constant of the fcc lattice we obtain

$$R_{\{111\}}(n,k) = (K/2) d_{\{111\}} = \frac{a}{4\sqrt{3}}(3n+k) \qquad n \geq k \qquad (C.40)$$

$$R_{\{100\}}(n,k) = N d_{\{100\}} = \frac{a}{4}(n+k) \qquad (C.41)$$

The distances between the center of a truncated cubic fcc(n, k) NP and its corners (**corner distances**) are given by $R_c(n, k)$ (referring to each of six ($n = k$) or 24 ($n > k$) corners at $\{1\,0\,0\}$ facets) where with (C.11) and $a$ denoting the lattice constant of the fcc lattice

$$R_c(n,k) = \frac{a}{4}\sqrt{3(k+n)^2 - 4kn} = \frac{a}{2}\sqrt{(2N-K)^2 + 2N^2} \qquad n \geq k \qquad (C.42)$$

The areas of each of the $\{1\,1\,1\}$ and $\{1\,0\,0\}$ facets of a truncated cubic fcc(n, k) NP, measured by corresponding corner atoms (**facet areas**), are given by $A_{\{111\}}(n, k)$ (referring to each of eight hexagonal (triangular) $\{1\,1\,1\}$ facets) and $A_{\{100\}}(n, k)$ (referring to each of six $\{1\,0\,0\}$ facets) where, with (C.2), (C.3), (C.11) and $a$ denoting the lattice constant of the fcc lattice

$$A_{(111)}(n,k) = \frac{\sqrt{3}}{8} a^2 k^2 = \frac{\sqrt{3}}{8} a^2 (3N-K)^2 \qquad (C.43)$$

$$A_{(100)}(n,k) = \frac{a^2}{4}\left[(n+k)^2 - 2k^2\right] = a^2\left[N^2 - \frac{1}{2}(3N-K)^2\right] \qquad (C.44)$$

## III. Conclusion

The present work gives a full account of the shape and structure of ideal nanoparticles (NPs) forming compact polyhedral sections of the ideal cubic lattice where simple, body centered, and face centered variants are considered. We have studied particles of $O_h$ symmetry which are confined by facets of densest and second densest monolayers of the lattice reflecting Miller index families {1 0 0}, {1 1 0} for sc and bcc as well as {1 1 1}, {1 0 0} for fcc. The structure evaluation identifies different types of generic NPs which serve for the definition of general polyhedral NPs. These can be classified according to integer valued structure parameters $N$, $M$ (, $K$) which are connected with particle diameters along corresponding facet normal directions reflecting {$h\,k\,l$} monolayer families of the underlying lattice. An alternative classification is based on parameters $n$, $m$ (, $k$) which can describe characteristic lengths of facet edges of the NPs where the parameters of the two classification are connected by linear transformations. Detailed structural properties of the general polyhedral NPs, such as shape, size, and surfaces, are discussed in analytical and numerical detail with visualization of characteristic examples.

Clearly, the present results deal only with ideal cubic NPs and cannot account for all possible structures of the most general metal nanoparticles observed, for example, by electron microscopy [a13]. Realistic NPs may exhibit very different shapes, including less compact particles, and symmetry, including local structural disorder and deviations from (incompatibility with) crystal lattice structure in their inner core. This can only be examined in case-by-case studies where exact quantitative results are difficult to obtain. However, the present results can be used in an approximate way to estimate typical particle sizes of metal NPs as well as for a repository of possible structures of compact NPs with internal cubic lattice.

# V. Supplementary Information

## S.1. Symmetry Centers

The cubic NPs, sc($n$, $m$), bcc($n$, $m$), and fcc($n$, $m$) of $O_h$ symmetry contain atoms or high symmetry voids at their center depending on the lattice type and on the $N$, $M$, $K$, $n$, $m$, $k$ as described in the following.

The **simple cubic** lattice offers two different centers of $O_h$ symmetry, an atom site and a high symmetry void as shown in Fig. S.1.

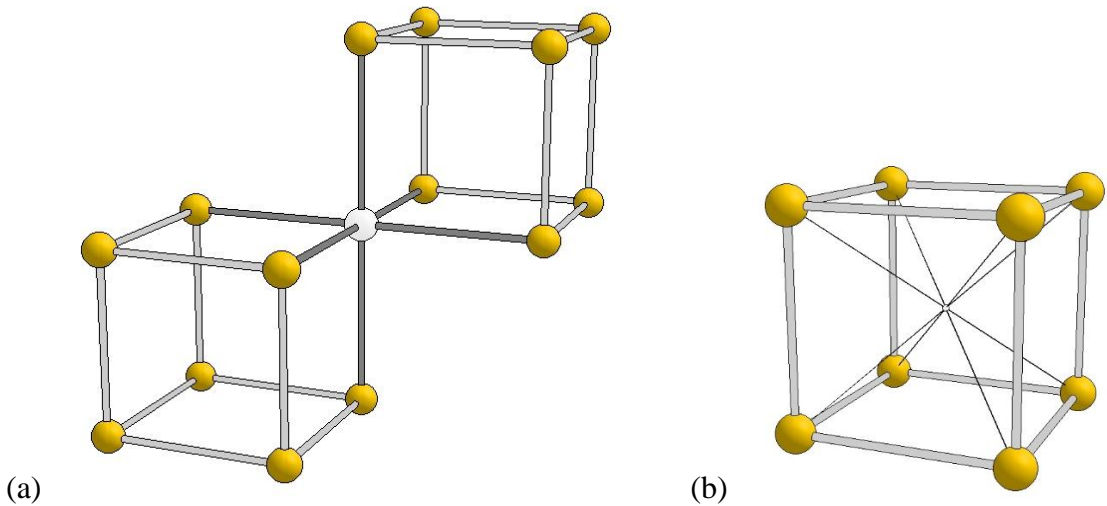

(a)           (b)

**Figure S.1.** $O_h$ symmetry centers of the simple cubic lattice, (a) center at atom site, (b) center at high symmetry void. The centers are emphasized by white balls and connected with their nearest neighbor atoms by dark sticks and black lines, respectively.

This discriminates between two types of sc[$N$, $M$], sc($n$, $m$) NPs, those about an atom center and those about a high symmetry void, as discussed in Sec. II.A.1 and spelled out in the following table.

| Center type | $N$ | $M$ | $n$ | $m$ |
|---|---|---|---|---|
| Atom | even | any | even | any |
| Void | odd | any | odd | any |

$n = 2M - N$ ,    $m = N - M$

The **body centered cubic** lattice offers only one center of $O_h$ symmetry which coincides with an atom site as shown in Fig. S.2.

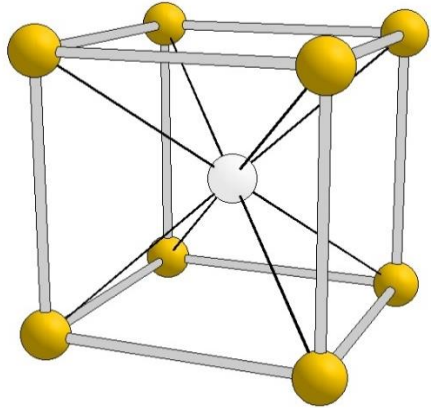

**Figure S.2.** $O_h$ symmetry center of the body centered cubic lattice at atom site. The center is emphasized by a white ball and connected with its nearest neighbor atoms by black lines.

This allows for only one type of bcc[$N$, $M$], bcc($n$, $m$) NPs, about an atom center as discussed in Sec. II.B.1 where $n = 2M - N$, $m = N - M$.

The **face centered cubic** lattice offers two different centers of $O_h$ symmetry, an atom site and a high symmetry void as shown in Fig. S.3.

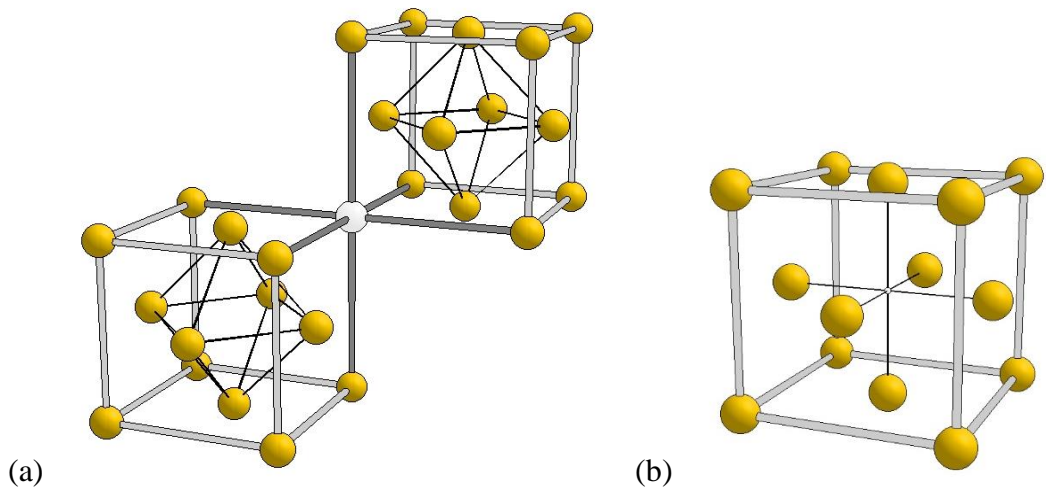

(a)                                        (b)

**Figure S.3.** $O_h$ symmetry centers of the face centered cubic lattice, (a) center at atom site, (b) center at high symmetry void. The centers are emphasized by white balls and connected with their nearest neighbor atoms by dark sticks and black lines, respectively.

This discriminates between two types of octahedral and cubic fcc[$N$, $K$], fcc($n$, $k$) NPs, about an atom center and about a high symmetry void, as discussed in Sec. II.C.1 and spelled out in the following table.

| Center type | $N$ | $K$ | $n$ | $k$ |
|---|---|---|---|---|
| Atom | even | even | even | even |
| Atom | odd | even | odd | odd |
| Void | even | odd | odd | odd |
| Void | odd | odd | even | even |

$n = K - N$ , $\quad k = 3N - K$

## S.2. Nanoparticle Peeling and Dressing

Taking off all atoms of the outermost {$hkl$} facet layer of a general cubic NP, denoted by structure parameters [$N$, $M$], ($n$, $m$) or [$N$, $K$], ($n$, $k$), (**{$hkl$} peeling**) leads to a smaller NP, denoted by [$N'$, $M'$], ($n'$, $m'$) or [$N'$, $K'$], ($n'$, $k'$) whose structure can be evaluated analytically. Further, adding all atoms from possible nearest outermost {$hkl$} monolayer facets (**{$hkl$} dressing**) leads to a larger NP and can also be evaluated analytically as discussed below. Successive peeling steps can reduce the cubic NP to a primitive (0, 0), (1, 0), or (0, 1) NP. This provides a detailed analysis of the internal structure of the NP.

### S.2.1. Cubo-rhombic sc Nanoparticles

Peeling and dressing a cubo-rhombic sc[$N$, $M$], sc($n$, $m$) NP of simple cubic lattice structure can be expressed by transformed structure parameters $N'$, $M'$, $n'$, $m'$ which depend on the initial parameters $M$, $N$ and are listed in the following tables. For the sake of brevity the prefix "sc" has been omitted in all NP notations.

- **{1 0 0} peeling /dressing**

| Nanoparticle | Peeled sc NP<br>[$N'$, $M'$] / ($n'$, $m'$) | Dressed sc NP<br>[$N'$, $M'$] / ($n'$, $m'$) |
|---|---|---|
| Cubo-rhombic<br>$M + 2 \leq N \leq 2M - 2$ | [$N - 2$, $M$] / ($n + 2$, $m - 2$) | [$N + 2$, $M$] / ($n - 2$, $m + 2$) |
| Generic cubic<br>$N = M$, $m = 0$ | [$N - 2$, $M - 2$] / ($n - 2$, 0) | [$N + 2$, $M$] / ($n - 2$, 2) |
| Cubo-rhombic<br>$N = M + 1$, $m = 1$ | [$N - 2$, $M - 1$] / ($n$, 0) | [$N + 2$, $M$] / ($n - 2$, 3) |
| Generic rhombic A<br>$N = 2M$, $n = 0$ | [$N - 2$, $M$] / (2, $m - 2$) | [$N$, $M$] / (0, $m$) |
| Generic rhombic B<br>$N = 2M - 1$, $n = 1$ | [$N - 2$, $M$] / (3, $m - 2$) | [$N + 2$, $M + 1$] / (1, $m + 1$) |

$n = 2M - N$, $m = N - M$, $N = n + 2m$, $M = n + m$, $M \leq N \leq 2M$

- **{1 1 0} peeling /dressing**

| Nanoparticle | Peeled sc NP<br>[$N'$, $M'$] / ($n'$, $m'$) | Dressed sc NP<br>[$N'$, $M'$] / ($n'$, $m'$) |
|---|---|---|
| Cubo-rhombic<br>$M \leq N \leq 2M - 2$ | [$N$, $M - 1$] / ($n - 2$, $m + 1$) | [$N$, $M + 1$] / ($n + 2$, $m - 1$) |
| Generic cubic<br>$N = M$, $m = 0$ | [$N$, $M - 1$] / ($n - 2$, 1) | [$N$, $M$] / ($n$, 0) |
| Generic rhombic A<br>$N = 2M$, $n = 0$ | [$N - 2$, $M - 1$] / (0, $m - 1$) | [$N$, $M + 1$] / (2, $m - 1$) |
| Generic rhombic B<br>$N = 2M - 1$, $n = 1$ | [$N - 2$, $M - 1$] / (1, $m - 1$) | [$N$, $M + 1$] / (3, $m - 1$) |

$n = 2M - N$, $m = N - M$, $N = n + 2m$, $M = n + m$, $M \leq N \leq 2M$

- **shell (joint {1 0 0} and {1 1 0}) peeling /dressing**

| Nanoparticle | Peeled sc NP<br>[$N'$, $M'$] / ($n'$, $m'$) | Dressed sc NP<br>[$N'$, $M'$] / ($n'$, $m'$) |
|---|---|---|
| Cubo-rhombic<br>$M + 2 \leq N \leq 2M - 1$ | [$N - 2$, $M - 1$] / ($n$, $m - 1$) | [$N + 2$, $M + 1$] / ($n$, $m + 1$) |
| Generic cubic<br>$N = M$, $m = 0$ | [$N - 2$, $M - 2$] / ($n - 2$, 0) | [$N + 2$, $M + 1$] / ($n$, 1) |
| Generic rhombic A<br>$N = 2M$, $n = 0$ | [$N - 2$, $M - 1$] / (0, $m - 1$) | [$N + 2$, $M + 1$] / (0, $m + 1$) |
| Generic rhombic B<br>$N = 2M - 1$, $n = 1$ | [$N - 2$, $M - 1$] / (1, $m - 1$) | [$N + 2$, $M + 1$] / (1, $m + 1$) |

$n = 2M - N$, $m = N - M$, $N = n + 2m$, $M = n + m$, $M \leq N \leq 2M$

## S.2.2. Cubo-rhombic bcc Nanoparticles

Peeling and dressing a cubo-rhombic bcc[$N$, $M$], bcc($n$, $m$) NP of body centered cubic lattice structure can be expressed by transformed structure parameters $N'$, $M'$, $n'$, $m'$ which depend on the initial parameters $M$, $N$ and are listed in the following tables. For the sake of brevity the prefix "bcc" has been omitted in all NP notations.

- **{1 0 0} peeling /dressing**

| Nanoparticle | Peeled bcc NP [$N'$, $M'$] / ($n'$, $m'$) | Dressed bcc NP [$N'$, $M'$] / ($n'$, $m'$) |
|---|---|---|
| Cubo-rhombic $M + 1 \leq N \leq 2M - 1$ | [$N - 1$, $M$] / ($n + 1$, $m - 1$) | [$N + 1$, $M$] / ($n - 1$, $m + 1$) |
| Generic cubic $N = M$, $m = 0$ | [$N - 1$, $M - 1$] / ($n - 1$, 0) | [$N + 1$, $M$] / ($n - 1$, 1) |
| Generic rhombic $N = 2M$, $n = 0$ | [$N - 1$, $M$] / (1, $m - 1$) | [$N$, $M$] / (0, $m$) |

$n = 2M - N$, $m = N - M$, $N = n + 2m$, $M = n + m$, $M \leq N \leq 2M$

- **{1 1 0} peeling /dressing**

| Nanoparticle | Peeled bcc NP [$N'$, $M'$] / ($n'$, $m'$) | Dressed bcc NP [$N'$, $M'$] / ($n'$, $m'$) |
|---|---|---|
| Cubo-rhombic $M + 1 \leq N \leq 2M - 2$ | [$N$, $M - 1$] / ($n - 2$, $m + 1$) | [$N$, $M + 1$] / ($n + 2$, $m - 1$) |
| Generic cubic $N = M$, $m = 0$ | [$N$, $M - 1$] / ($n - 2$, 1) | [$N$, $M$] / ($n$, 0) |
| Generic rhombic $N = 2M$, $n = 0$ | [$N - 2$, $M - 1$] / (0, $m - 1$) | [$N$, $M + 1$] / (2, $m - 1$) |

$n = 2M - N$, $m = N - M$, $N = n + 2m$, $M = n + m$, $M \leq N \leq 2M$

- **shell (joint {1 0 0} and {1 1 0}) peeling /dressing**

| Nanoparticle | Peeled bcc NP [$N'$, $M'$] / ($n'$, $m'$) | Dressed bcc NP [$N'$, $M'$] / ($n'$, $m'$) |
|---|---|---|
| Cubo-rhombic $M + 1 \leq N \leq 2M - 1$ | [$N - 1$, $M - 1$] / ($n - 1$, $m$) | [$N + 1$, $M + 1$] / ($n + 1$, $m$) |
| Generic cubic $N = M$, $m = 0$ | [$N - 1$, $M - 1$] / ($n - 1$, 0) | [$N + 1$, $M + 1$] / ($n + 1$, 0) |
| Generic rhombic $N = 2M$, $n = 0$ | [$N - 2$, $M - 1$] / (0, $m - 1$) | [$N + 1$, $M + 1$] / (1, $m$) |

$n = 2M - N$, $m = N - M$, $N = n + 2m$, $M = n + m$, $M \leq N \leq 2M$

### S.2.3. Cuboctahedral fcc Nanoparticles

Peeling and dressing a cuboctahedral fcc[$N$, $K$], fcc($n$, $k$) NP of face centerd cubic lattice structure can be expressed by transformed structure parameters $N'$, $K'$, $n'$, $k'$ which depend on the initial parameters $M$, $K$ and are listed in the following tables. For the sake of brevity the prefix "fcc" has been omitted in all NP notations.

As discussed in Sec. C.3, there are two types of face centered cubic NPs which are distinguished by their shape, truncated octahedral and truncated cubic fcc[$N$, $K$], fcc($n$, $k$), with $N$, $K$, $n$, $k$ determined by relations (C.12a) and (C.12b), respectively. Peeling and dressing induce transitions between the two types, cubic to octahedral for peeling and octahedral to cubic for dressing, but do not affect corresponding transformations $N$, $K$, $n$, $k$ to $N'$, $K'$, $n'$, $k'$.

- **{1 0 0} peeling /dressing**

| Nanoparticle | Peeled fcc NP $[N', K'] / (n', k')$ | Dressed fcc NP $[N', K'] / (n', k')$ |
|---|---|---|
| Cuboctahedral $N \leq K \leq 3N - 3$ | $[N -1, K] / (n + 1, k - 3)$ | $[N + 1, K] / (n - 1, k + 3)$ |
| Generic cubic A $K = 3N$, $k = 0$ | $[N - 1, K - 4] / (n - 3, 1)$ | $[N + 1, K] / (n - 1, 3)$ |
| Generic cubic B $K = 3N - 1$, $k = 1$ | $[N - 1, K - 2] / (n - 1, 0)$ | $[N + 1, K] / (n - 1, 4)$ |
| Generic cubic C $K = 3N - 2$, $k = 2$ | $[N - 1, K - 2] / (n - 1, 1)$ | $[N + 1, K] / (n - 1, 5)$ |
| Generic octahedral A $K = N$, $n = 0$ | $[N - 1, K] / (1, k - 3)$ | $[N, K] / (0, k)$ |
| Generic octahedral B $K = N + 1$, $n = 1$ | $[N - 1, K] / (2, k - 3)$ | $[N + 1, K] / (0, k + 3)$ |
| Generic cuboctahedral $K = 2N$, $n = k$ | $[N - 1, K] / (n + 1, k - 3)$ | $[N + 1, K] / (n - 1, k + 3)$ |

$n = K - N$,   $k = 3N - K$,   $2N = n + k$,   $2K = 3n + k$,   $N \leq K \leq 3N$

- **{1 1 1} peeling /dressing**

| Nanoparticle | Peeled fcc NP [N', K'] / (n', k') | Dressed fcc NP [N', K'] / (n', k') |
|---|---|---|
| Cuboctahedral $N + 2 \leq K \leq 3N - 2$ | [N, K - 2] / (n - 2, k + 2) | [N, K + 2] / (n + 2, k - 2) |
| Generic cubic A $K = 3N, \ k = 0$ | [N, K - 2] / (n - 2, 2) | [N, K] / (n, 0) |
| Generic cubic B $K = 3N - 1, \ k = 1$ | [N, K - 2] / (n - 2, 3) | [N, K] / (n, 1) |
| Generic cubic C $K = 3N - 2, \ k = 2$ | [N, K - 2] / (n - 2, 4) | [N, K + 2] / (n + 2, 0) |
| Generic octahedral A $K = N, \ n = 0$ | [N - 2, K - 2] / (0, k - 4) | [N, K + 2] / (2, k - 2) |
| Generic octahedral B $K = N + 1, \ n = 1$ | [N - 1, K - 2] / (0, k - 1) | [N, K + 2] / (3, k - 2) |
| Generic cuboctahedral $K = 2N, \ n = k$ | [N, K - 2] / (n - 2, k + 2) | [N, K + 2] / (n + 2, k - 2) |

$n = K - N, \quad k = 3N - K, \quad 2N = n + k, \quad 2K = 3n + k, \quad N \leq K \leq 3N$

- **shell (joint {1 0 0} and {1 1 1}) peeling /dressing**

| Nanoparticle | Peeled fcc NP [N', K'] / (n', k') | Dressed fcc NP [N', K'] / (n', k') |
|---|---|---|
| Cuboctahedral $N + 1 \leq K \leq 3N - 1$ | [N - 1, K - 2] / (n - 1, k - 1) | [N + 1, K + 2] / (n + 1, k + 1) |
| Generic cubic A $K = 3N, \ k = 0$ | [N - 1, K - 4] / (n - 3, 1) | [N + 1, K + 2] / (n + 1, 1) |
| Generic cubic B $K = 3N - 1, \ k = 1$ | [N - 1, K - 2] / (n - 1, 0) | [N + 1, K + 2] / (n + 1, 2) |
| Generic cubic C $K = 3N - 2, \ k = 2$ | [N - 1, K - 2] / (n - 1, 1) | [N + 1, K + 2] / (n + 1, 3) |
| Generic octahedral A $K = N, \ n = 0$ | [N - 2, K - 2] / (0, k - 4) | [N + 1, K + 2] / (1, k + 1) |
| Generic octahedral B $K = N + 1, \ n = 1$ | [N - 1, K - 2] / (0, k - 1) | [N + 1, K + 2] / (2, k + 1) |
| Generic cuboctahedral $K = 2N, \ n = k$ | [N - 1, K - 2] / (n - 1, k - 1) | [N + 1, K + 2] / (n + 1, k + 1) |

$n = K - N, \quad k = 3N - K, \quad 2N = n + k, \quad 2K = 3n + k, \quad N \leq K \leq 3N$

## S.3. Numerical Tables

### S.3.1. Cubo-rhombic sc Nanoparticles

**Table 1.1.** Notation of a non-generic cubo-rhombic NP sc[$N$, $M$], sc($n$, $m$) for given $n$, $m$ according to $N = n + 2m$, $M = n + m$. Structure parameters $N$, $M$ describe the two generic envelope NPs, cubic sc[$N$, 0] and rhombic A sc(0, $M$] ($n$ even) or rhombic B sc(1, $M$) ($n$ odd) by polyhedral diameters, see Sec. A.3. In contrast, $n$, $m$ characterize the deviation of the cubo-rhombic NP sc[$N$, $M$] from its generic envelope NPs.

List of [$N$, $M$]

```
m\n |    0       1       2       3       4       5       6       7       8       9      10      11
  0 | [ 0, 0][ 1, 1][ 2, 2][ 3, 3][ 4, 4][ 5, 5][ 6, 6][ 7, 7][ 8, 8][ 9, 9][10,10][11,11]
  1 | [ 2, 1][ 3, 2][ 4, 3][ 5, 4][ 6, 5][ 7, 6][ 8, 7][ 9, 8][10, 9][11,10][12,11][13,12]
  2 | [ 4, 2][ 5, 3][ 6, 4][ 7, 5][ 8, 6][ 9, 7][10, 8][11, 9][12,10][13,11][14,12][15,13]
  3 | [ 6, 3][ 7, 4][ 8, 5][ 9, 6][10, 7][11, 8][12, 9][13,10][14,11][15,12][16,13][17,14]
  4 | [ 8, 4][ 9, 5][10, 6][11, 7][12, 8][13, 9][14,10][15,11][16,12][17,13][18,14][19,15]
  5 | [10, 5][11, 6][12, 7][13, 8][14, 9][15,10][16,11][17,12][18,13][19,14][20,15][21,16]
  6 | [12, 6][13, 7][14, 8][15, 9][16,10][17,11][18,12][19,13][20,14][21,15][22,16][23,17]
  7 | [14, 7][15, 8][16, 9][17,10][18,11][19,12][20,13][21,14][22,15][23,16][24,17][25,18]
  8 | [16, 8][17, 9][18,10][19,11][20,12][21,13][22,14][23,15][24,16][25,17][26,18][27,19]
  9 | [18, 9][19,10][20,11][21,12][22,13][23,14][24,15][25,16][26,17][27,18][28,19][29,20]
 10 | [20,10][21,11][22,12][23,13][24,14][25,15][26,16][27,17][28,18][29,19][30,20][31,21]
 11 | [22,11][23,12][24,13][25,14][26,15][27,16][28,17][29,18][30,19][31,20][32,21][33,22]
```

**Table 1.2.** Number of atoms inside the volume of an sc($n$, $m$) NP, $N_{\text{vol}}(n, m)$, (volume count) as defined by (A.17 - 20).

List of $N_{\text{vol}}(n, m)$

```
m\n  |    0      1      2      3      4      5      6      7      8      9     10     11
  0  |    1      8     27     64    125    216    343    512    729   1000   1331   1728
  1  |    7     32     81    160    275    432    637    896   1215   1600   2057   2592
  2  |   33     88    179    312    493    728   1023   1384   1817   2328   2923   3608
  3  |   87    184    329    528    787   1112   1509   1984   2543   3192   3937   4784
  4  |  185    336    547    824   1173   1600   2111   2712   3409   4208   5115   6136
  5  |  335    552    841   1208   1659   2200   2837   3576   4423   5384   6465   7672
  6  |  553    848   1227   1696   2261   2928   3703   4592   5601   6736   8003   9408
  7  |  847   1232   1713   2296   2987   3792   4717   5768   6951   8272   9737  11352
  8  | 1233   1720   2315   3024   3853   4808   5895   7120   8489  10008  11683  13520
  9  | 1719   2320   3041   3888   4867   5984   7245   8656  10223  11952  13849  15920
 10  | 2321   3048   3907   4904   6045   7336   8783  10392  12169  14120  16251  18568
 11  | 3047   3912   4921   6080   7395   8872  10517  12336  14335  16520  18897  21472
```

**Table 1.3.** Number of atoms on the outermost facets of an sc($n$, $m$) NP, $N_{shell}(n, m)$, (facet count) as defined by (A.21 - 24).

List of $N_{shell}(n, m)$

| m\n | 0 | 1 | 2 | 3 | 4 | 5 | 6 | 7 | 8 | 9 | 10 | 11 |
|---|---|---|---|---|---|---|---|---|---|---|---|---|
| 0 | -- | 8 | 26 | 56 | 98 | 152 | 218 | 296 | 386 | 488 | 602 | 728 |
| 1 | 6 | 24 | 54 | 96 | 150 | 216 | 294 | 384 | 486 | 600 | 726 | 864 |
| 2 | 26 | 56 | 98 | 152 | 218 | 296 | 386 | 488 | 602 | 728 | 866 | 1016 |
| 3 | 54 | 96 | 150 | 216 | 294 | 384 | 486 | 600 | 726 | 864 | 1014 | 1176 |
| 4 | 98 | 152 | 218 | 296 | 386 | 488 | 602 | 728 | 866 | 1016 | 1178 | 1352 |
| 5 | 150 | 216 | 294 | 384 | 486 | 600 | 726 | 864 | 1014 | 1176 | 1350 | 1536 |
| 6 | 218 | 296 | 386 | 488 | 602 | 728 | 866 | 1016 | 1178 | 1352 | 1538 | 1736 |
| 7 | 294 | 384 | 486 | 600 | 726 | 864 | 1014 | 1176 | 1350 | 1536 | 1734 | 1944 |
| 8 | 386 | 488 | 602 | 728 | 866 | 1016 | 1178 | 1352 | 1538 | 1736 | 1946 | 2168 |
| 9 | 486 | 600 | 726 | 864 | 1014 | 1176 | 1350 | 1536 | 1734 | 1944 | 2166 | 2400 |
| 10 | 602 | 728 | 866 | 1016 | 1178 | 1352 | 1538 | 1736 | 1946 | 2168 | 2402 | 2648 |
| 11 | 726 | 864 | 1014 | 1176 | 1350 | 1536 | 1734 | 1944 | 2166 | 2400 | 2646 | 2904 |

**Table 1.4.** Corner distances, $R_{c1}(n, m)$, defining the distances from the center of a simple cubic sc($n$, $m$) NP to its 24 ($n > 0$) and 6 ($n = 0$) corners, respectively, at {1 0 0} facets and $R_{c2}(n, m)$, defining the distances to its 8 ($m$ even) and 24 ($m$ odd) corners, respectively, at {1 1 0} (and 8 possible {1 1 1}) facets. The distances are defined by (A.29), (A.30) and all radii are normalized by the lattice constant $a$ of the sc lattice.

List of $R_{c1}(n, m) / a$

| m\n | 0 | 1 | 2 | 3 | 4 | 5 | 6 | 7 | 8 | 9 | 10 | 11 |
|---|---|---|---|---|---|---|---|---|---|---|---|---|
| 0 | 0.000 | 0.866 | 1.732 | 2.598 | 3.464 | 4.330 | 5.196 | 6.062 | 6.928 | 7.794 | 8.660 | 9.526 |
| 1 | 1.000 | 1.658 | 2.449 | 3.279 | 4.123 | 4.975 | 5.831 | 6.690 | 7.550 | 8.411 | 9.274 | 10.137 |
| 2 | 2.000 | 2.598 | 3.317 | 4.093 | 4.899 | 5.723 | 6.557 | 7.399 | 8.246 | 9.097 | 9.950 | 10.805 |
| 3 | 3.000 | 3.571 | 4.243 | 4.975 | 5.745 | 6.538 | 7.348 | 8.170 | 9.000 | 9.836 | 10.677 | 11.522 |
| 4 | 4.000 | 4.555 | 5.196 | 5.895 | 6.633 | 7.399 | 8.185 | 8.986 | 9.798 | 10.618 | 11.446 | 12.278 |
| 5 | 5.000 | 5.545 | 6.164 | 6.837 | 7.550 | 8.292 | 9.055 | 9.836 | 10.630 | 11.435 | 12.247 | 13.067 |
| 6 | 6.000 | 6.538 | 7.141 | 7.794 | 8.485 | 9.206 | 9.950 | 10.712 | 11.489 | 12.278 | 13.077 | 13.883 |
| 7 | 7.000 | 7.533 | 8.124 | 8.761 | 9.434 | 10.137 | 10.863 | 11.608 | 12.369 | 13.143 | 13.928 | 14.722 |
| 8 | 8.000 | 8.529 | 9.110 | 9.734 | 10.392 | 11.079 | 11.790 | 12.520 | 13.266 | 14.027 | 14.799 | 15.580 |
| 9 | 9.000 | 9.526 | 10.100 | 10.712 | 11.358 | 12.031 | 12.728 | 13.444 | 14.177 | 14.925 | 15.684 | 16.454 |
| 10 | 10.000 | 10.524 | 11.091 | 11.694 | 12.329 | 12.990 | 13.675 | 14.379 | 15.100 | 15.835 | 16.583 | 17.342 |
| 11 | 11.000 | 11.522 | 12.083 | 12.679 | 13.304 | 13.955 | 14.629 | 15.322 | 16.031 | 16.756 | 17.493 | 18.241 |

List of $R_{c2}(n, m) / a$

| m\n | 0 | 1 | 2 | 3 | 4 | 5 | 6 | 7 | 8 | 9 | 10 | 11 |
|---|---|---|---|---|---|---|---|---|---|---|---|---|
| 0 | 0.000 | 0.866 | 1.732 | 2.598 | 3.464 | 4.330 | 5.196 | 6.062 | 6.928 | 7.794 | 8.660 | 9.526 |
| 1 | 1.000 | 1.658 | 2.449 | 3.279 | 4.123 | 4.975 | 5.831 | 6.690 | 7.550 | 8.411 | 9.274 | 10.137 |
| 2 | 1.732 | 2.598 | 3.464 | 4.330 | 5.196 | 6.062 | 6.928 | 7.794 | 8.660 | 9.526 | 10.392 | 11.258 |
| 3 | 2.449 | 3.279 | 4.123 | 4.975 | 5.831 | 6.690 | 7.550 | 8.411 | 9.274 | 10.137 | 11.000 | 11.864 |
| 4 | 3.464 | 4.330 | 5.196 | 6.062 | 6.928 | 7.794 | 8.660 | 9.526 | 10.392 | 11.258 | 12.124 | 12.990 |
| 5 | 4.123 | 4.975 | 5.831 | 6.690 | 7.550 | 8.411 | 9.274 | 10.137 | 11.000 | 11.864 | 12.728 | 13.592 |
| 6 | 5.196 | 6.062 | 6.928 | 7.794 | 8.660 | 9.526 | 10.392 | 11.258 | 12.124 | 12.990 | 13.856 | 14.722 |
| 7 | 5.831 | 6.690 | 7.550 | 8.411 | 9.274 | 10.137 | 11.000 | 11.864 | 12.728 | 13.592 | 14.457 | 15.322 |
| 8 | 6.928 | 7.794 | 8.660 | 9.526 | 10.392 | 11.258 | 12.124 | 12.990 | 13.856 | 14.722 | 15.588 | 16.454 |
| 9 | 7.550 | 8.411 | 9.274 | 10.137 | 11.000 | 11.864 | 12.728 | 13.592 | 14.457 | 15.322 | 16.186 | 17.051 |
| 10 | 8.660 | 9.526 | 10.392 | 11.258 | 12.124 | 12.990 | 13.856 | 14.722 | 15.588 | 16.454 | 17.321 | 18.187 |
| 11 | 9.274 | 10.137 | 11.000 | 11.864 | 12.728 | 13.592 | 14.457 | 15.322 | 16.186 | 17.051 | 17.916 | 18.782 |

## S.3.2. Cubo-rhombic bcc Nanoparticles

**Table 2.1.** Notation of a non-generic cubo-rhombic NP bcc[$N$, $M$], bcc($n$, $m$) for given $n$, $m$ according to $N = n + 2m$, $M = n + m$. Structure parameters $N$, $M$ describe the two generic envelope NPs, cubic sc[$N$, 0) and rhombic sc(0, $M$] by polyhedral diameters, see Sec. B.3. In contrast, $n$, $m$ characterize the deviation of the cubo-rhombic NP bcc[$N$, $M$] from its generic envelope NPs.

List of $N : M$

| m\n | 0 | 1 | 2 | 3 | 4 | 5 | 6 | 7 | 8 | 9 | 10 | 11 |
|---|---|---|---|---|---|---|---|---|---|---|---|---|
| 0 | [ 0, 0] | [ 1, 1] | [ 2, 2] | [ 3, 3] | [ 4, 4] | [ 5, 5] | [ 6, 6] | [ 7, 7] | [ 8, 8] | [ 9, 9] | [10,10] | [11,11] |
| 1 | [ 2, 1] | [ 3, 2] | [ 4, 3] | [ 5, 4] | [ 6, 5] | [ 7, 6] | [ 8, 7] | [ 9, 8] | [10, 9] | [11,10] | [12,11] | [13,12] |
| 2 | [ 4, 2] | [ 5, 3] | [ 6, 4] | [ 7, 5] | [ 8, 6] | [ 9, 7] | [10, 8] | [11, 9] | [12,10] | [13,11] | [14,12] | [15,13] |
| 3 | [ 6, 3] | [ 7, 4] | [ 8, 5] | [ 9, 6] | [10, 7] | [11, 8] | [12, 9] | [13,10] | [14,11] | [15,12] | [16,13] | [17,14] |
| 4 | [ 8, 4] | [ 9, 5] | [10, 6] | [11, 7] | [12, 8] | [13, 9] | [14,10] | [15,11] | [16,12] | [17,13] | [18,14] | [19,15] |
| 5 | [10, 5] | [11, 6] | [12, 7] | [13, 8] | [14, 9] | [15,10] | [16,11] | [17,12] | [18,13] | [19,14] | [20,15] | [21,16] |
| 6 | [12, 6] | [13, 7] | [14, 8] | [15, 9] | [16,10] | [17,11] | [18,12] | [19,13] | [20,14] | [21,15] | [22,16] | [23,17] |
| 7 | [14, 7] | [15, 8] | [16, 9] | [17,10] | [18,11] | [19,12] | [20,13] | [21,14] | [22,15] | [23,16] | [24,17] | [25,18] |
| 8 | [16, 8] | [17, 9] | [18,10] | [19,11] | [20,12] | [21,13] | [22,14] | [23,15] | [24,16] | [25,17] | [26,18] | [27,19] |
| 9 | [18, 9] | [19,10] | [20,11] | [21,12] | [22,13] | [23,14] | [24,15] | [25,16] | [26,17] | [27,18] | [28,19] | [29,20] |
| 10 | [20,10] | [21,11] | [22,12] | [23,13] | [24,14] | [25,15] | [26,16] | [27,17] | [28,18] | [29,19] | [30,20] | [31,21] |
| 11 | [22,11] | [23,12] | [24,13] | [25,14] | [26,15] | [27,16] | [28,17] | [29,18] | [30,19] | [31,20] | [32,21] | [33,22] |

**Table 2.2.** Number of atoms inside the volume of a bcc($n$, $m$) NP, $N_{\text{vol}}(n, m)$, (volume count) as defined by (B.12 - 14).

List of $N_{\text{vol}}(n, m)$

| m\n | 0 | 1 | 2 | 3 | 4 | 5 | 6 | 7 | 8 | 9 | 10 | 11 |
|---|---|---|---|---|---|---|---|---|---|---|---|---|
| 0 | 1 | 9 | 35 | 91 | 189 | 341 | 559 | 855 | 1241 | 1729 | 2331 | 3059 |
| 1 | 15 | 59 | 145 | 285 | 491 | 775 | 1149 | 1625 | 2215 | 2931 | 3785 | 4789 |
| 2 | 65 | 169 | 339 | 587 | 925 | 1365 | 1919 | 2599 | 3417 | 4385 | 5515 | 6819 |
| 3 | 175 | 363 | 641 | 1021 | 1515 | 2135 | 2893 | 3801 | 4871 | 6115 | 7545 | 9173 |
| 4 | 369 | 665 | 1075 | 1611 | 2285 | 3109 | 4095 | 5255 | 6601 | 8145 | 9899 | 11875 |
| 5 | 671 | 1099 | 1665 | 2381 | 3259 | 4311 | 5549 | 6985 | 8631 | 10499 | 12601 | 14949 |
| 6 | 1105 | 1689 | 2435 | 3355 | 4461 | 5765 | 7279 | 9015 | 10985 | 13201 | 15675 | 18419 |
| 7 | 1695 | 2459 | 3409 | 4557 | 5915 | 7495 | 9309 | 11369 | 13687 | 16275 | 19145 | 22309 |
| 8 | 2465 | 3433 | 4611 | 6011 | 7645 | 9525 | 11663 | 14071 | 16761 | 19745 | 23035 | 26643 |
| 9 | 3439 | 4635 | 6065 | 7741 | 9675 | 11879 | 14365 | 17145 | 20231 | 23635 | 27369 | 31445 |
| 10 | 4641 | 6089 | 7795 | 9771 | 12029 | 14581 | 17439 | 20615 | 24121 | 27969 | 32171 | 36739 |
| 11 | 6095 | 7819 | 9825 | 12125 | 14731 | 17655 | 20909 | 24505 | 28455 | 32771 | 37465 | 42549 |

**Table 2.3.** Number of atoms on the outermost facets of a bcc($n$, $m$) NP, $N_{\text{shell}}(n, m)$, (facet count) as defined by (B.15 - 17).

List of $N_{\text{shell}}(n, m)$

| m\n | 0 | 1 | 2 | 3 | 4 | 5 | 6 | 7 | 8 | 9 | 10 | 11 |
|---|---|---|---|---|---|---|---|---|---|---|---|---|
| 0 | -- | 8 | 26 | 56 | 98 | 152 | 218 | 296 | 386 | 488 | 602 | 728 |
| 1 | 14 | 44 | 86 | 140 | 206 | 284 | 374 | 476 | 590 | 716 | 854 | 1004 |
| 2 | 50 | 104 | 170 | 248 | 338 | 440 | 554 | 680 | 818 | 968 | 1130 | 1304 |
| 3 | 110 | 188 | 278 | 380 | 494 | 620 | 758 | 908 | 1070 | 1244 | 1430 | 1628 |
| 4 | 194 | 296 | 410 | 536 | 674 | 824 | 986 | 1160 | 1346 | 1544 | 1754 | 1976 |
| 5 | 302 | 428 | 566 | 716 | 878 | 1052 | 1238 | 1436 | 1646 | 1868 | 2102 | 2348 |
| 6 | 434 | 584 | 746 | 920 | 1106 | 1304 | 1514 | 1736 | 1970 | 2216 | 2474 | 2744 |
| 7 | 590 | 764 | 950 | 1148 | 1358 | 1580 | 1814 | 2060 | 2318 | 2588 | 2870 | 3164 |
| 8 | 770 | 968 | 1178 | 1400 | 1634 | 1880 | 2138 | 2408 | 2690 | 2984 | 3290 | 3608 |
| 9 | 974 | 1196 | 1430 | 1676 | 1934 | 2204 | 2486 | 2780 | 3086 | 3404 | 3734 | 4076 |
| 10 | 1202 | 1448 | 1706 | 1976 | 2258 | 2552 | 2858 | 3176 | 3506 | 3848 | 4202 | 4568 |
| 11 | 1454 | 1724 | 2006 | 2300 | 2606 | 2924 | 3254 | 3596 | 3950 | 4316 | 4694 | 5084 |

**Table 2.4.** Corner distances, $R_{c1}(n, m)$, defining the distances from the center of a body centered cubic bcc($n$, $m$) NP to its 24 ($n > 0$) and 6 ($n = 0$) corners, respectively, at {1 0 0} facets and $R_{c2}(n, m)$, defining the distances to its 8 corners at {1 1 0} facets. The distances are defined by (B.22), (B.23) and all radii are normalized by the lattice constant $a$ of the sc lattice.

List of $R_{c1}(n, m) / a$

| m\n | 0 | 1 | 2 | 3 | 4 | 5 | 6 | 7 | 8 | 9 | 10 | 11 |
|---|---|---|---|---|---|---|---|---|---|---|---|---|
| 0 | 0.000 | 0.866 | 1.732 | 2.598 | 3.464 | 4.330 | 5.196 | 6.062 | 6.928 | 7.794 | 8.660 | 9.526 |
| 1 | 1.000 | 1.658 | 2.449 | 3.279 | 4.123 | 4.975 | 5.831 | 6.690 | 7.550 | 8.411 | 9.274 | 10.137 |
| 2 | 2.000 | 2.598 | 3.317 | 4.093 | 4.899 | 5.723 | 6.557 | 7.399 | 8.246 | 9.097 | 9.950 | 10.805 |
| 3 | 3.000 | 3.571 | 4.243 | 4.975 | 5.745 | 6.538 | 7.348 | 8.170 | 9.000 | 9.836 | 10.677 | 11.522 |
| 4 | 4.000 | 4.555 | 5.196 | 5.895 | 6.633 | 7.399 | 8.185 | 8.986 | 9.798 | 10.618 | 11.446 | 12.278 |
| 5 | 5.000 | 5.545 | 6.164 | 6.837 | 7.550 | 8.292 | 9.055 | 9.836 | 10.630 | 11.435 | 12.247 | 13.067 |
| 6 | 6.000 | 6.538 | 7.141 | 7.794 | 8.485 | 9.206 | 9.950 | 10.712 | 11.489 | 12.278 | 13.077 | 13.883 |
| 7 | 7.000 | 7.533 | 8.124 | 8.761 | 9.434 | 10.137 | 10.863 | 11.608 | 12.369 | 13.143 | 13.928 | 14.722 |
| 8 | 8.000 | 8.529 | 9.110 | 9.734 | 10.392 | 11.079 | 11.790 | 12.520 | 13.266 | 14.027 | 14.799 | 15.580 |
| 9 | 9.000 | 9.526 | 10.100 | 10.712 | 11.358 | 12.031 | 12.728 | 13.444 | 14.177 | 14.925 | 15.684 | 16.454 |
| 10 | 10.000 | 10.524 | 11.091 | 11.694 | 12.329 | 12.990 | 13.675 | 14.379 | 15.100 | 15.835 | 16.583 | 17.342 |
| 11 | 11.000 | 11.522 | 12.083 | 12.679 | 13.304 | 13.955 | 14.629 | 15.322 | 16.031 | 16.756 | 17.493 | 18.241 |

List of $R_{c2}(n, m) / a$

| m\n | 0 | 1 | 2 | 3 | 4 | 5 | 6 | 7 | 8 | 9 | 10 | 11 |
|---|---|---|---|---|---|---|---|---|---|---|---|---|
| 0 | 0.000 | 0.866 | 1.732 | 2.598 | 3.464 | 4.330 | 5.196 | 6.062 | 6.928 | 7.794 | 8.660 | 9.526 |
| 1 | 0.866 | 1.732 | 2.598 | 3.464 | 4.330 | 5.196 | 6.062 | 6.928 | 7.794 | 8.660 | 9.526 | 10.392 |
| 2 | 1.732 | 2.598 | 3.464 | 4.330 | 5.196 | 6.062 | 6.928 | 7.794 | 8.660 | 9.526 | 10.392 | 11.258 |
| 3 | 2.598 | 3.464 | 4.330 | 5.196 | 6.062 | 6.928 | 7.794 | 8.660 | 9.526 | 10.392 | 11.258 | 12.124 |
| 4 | 3.464 | 4.330 | 5.196 | 6.062 | 6.928 | 7.794 | 8.660 | 9.526 | 10.392 | 11.258 | 12.124 | 12.990 |
| 5 | 4.330 | 5.196 | 6.062 | 6.928 | 7.794 | 8.660 | 9.526 | 10.392 | 11.258 | 12.124 | 12.990 | 13.856 |
| 6 | 5.196 | 6.062 | 6.928 | 7.794 | 8.660 | 9.526 | 10.392 | 11.258 | 12.124 | 12.990 | 13.856 | 14.722 |
| 7 | 6.062 | 6.928 | 7.794 | 8.660 | 9.526 | 10.392 | 11.258 | 12.124 | 12.990 | 13.856 | 14.722 | 15.588 |
| 8 | 6.928 | 7.794 | 8.660 | 9.526 | 10.392 | 11.258 | 12.124 | 12.990 | 13.856 | 14.722 | 15.588 | 16.454 |
| 9 | 7.794 | 8.660 | 9.526 | 10.392 | 11.258 | 12.124 | 12.990 | 13.856 | 14.722 | 15.588 | 16.454 | 17.321 |
| 10 | 8.660 | 9.526 | 10.392 | 11.258 | 12.124 | 12.990 | 13.856 | 14.722 | 15.588 | 16.454 | 17.321 | 18.187 |
| 11 | 9.526 | 10.392 | 11.258 | 12.124 | 12.990 | 13.856 | 14.722 | 15.588 | 16.454 | 17.321 | 18.187 | 19.053 |

## S.3.3a. Cuboctahedral fcc Nanoparticles, Truncated Octahedral Type

**Table 3.1.** Notation of a non-generic cuboctahedral NP fcc[$N$, $K$], fcc($n$, $k$) for given $n$, $k$ according to $2N = n + k$, $2K = 3n + k$. Structure parameters $N$, $K$ describe the two generic envelope NPs, cubic fcc[$N$, 0], fcc[$N$, 1] and octahedral sc(0, $K$), sc(1, $K$) by polyhedral diameters, see Sec. C.3. In contrast, $n$, $k$ characterize the deviation of the cuboctahedral NP fcc[$N$, $M$] from its generic envelope NPs.

For fcc[$N$, $K$] NPs of the octahedral type, i.e. if $N \leq K \leq 2N$, $k \geq n$, it is meaningful to use an auxiliary index $k'$ with $k' = (k - n)/2$ resulting in a notation fcc($n$, $k'$)$_o$, see Sec. C.3, and yielding

$$[N, K]: \quad N = n + k', \quad K = 2n + k'$$

List of [$N$, $K$]

```
k'\n |    0        1        2        3        4        5        6        7        8        9       10       11
   0 | [ 0, 0] [ 1, 2] [ 2, 4] [ 3, 6] [ 4, 8] [ 5,10] [ 6,12] [ 7,14] [ 8,16] [ 9,18] [10,20] [11,22]
   1 | [ 1, 1] [ 2, 3] [ 3, 5] [ 4, 7] [ 5, 9] [ 6,11] [ 7,13] [ 8,15] [ 9,17] [10,19] [11,21] [12,23]
   2 | [ 2, 2] [ 3, 4] [ 4, 6] [ 5, 8] [ 6,10] [ 7,12] [ 8,14] [ 9,16] [10,18] [11,20] [12,22] [13,24]
   3 | [ 3, 3] [ 4, 5] [ 5, 7] [ 6, 9] [ 7,11] [ 8,13] [ 9,15] [10,17] [11,19] [12,21] [13,23] [14,25]
   4 | [ 4, 4] [ 5, 6] [ 6, 8] [ 7,10] [ 8,12] [ 9,14] [10,16] [11,18] [12,20] [13,22] [14,24] [15,26]
   5 | [ 5, 5] [ 6, 7] [ 7, 9] [ 8,11] [ 9,13] [10,15] [11,17] [12,19] [13,21] [14,23] [15,25] [16,27]
   6 | [ 6, 6] [ 7, 8] [ 8,10] [ 9,12] [10,14] [11,16] [12,18] [13,20] [14,22] [15,24] [16,26] [17,28]
   7 | [ 7, 7] [ 8, 9] [ 9,11] [10,13] [11,15] [12,17] [13,19] [14,21] [15,23] [16,25] [17,27] [18,29]
   8 | [ 8, 8] [ 9,10] [10,12] [11,14] [12,16] [13,18] [14,20] [15,22] [16,24] [17,26] [18,28] [19,30]
   9 | [ 9, 9] [10,11] [11,13] [12,15] [13,17] [14,19] [15,21] [16,23] [17,25] [18,27] [19,29] [20,31]
  10 | [10,10] [11,12] [12,14] [13,16] [14,18] [15,20] [16,22] [17,24] [18,26] [19,28] [20,30] [21,32]
  11 | [11,11] [12,13] [13,15] [14,17] [15,19] [16,21] [17,23] [18,25] [19,27] [20,29] [21,31] [22,33]
```

**Table 3.2.** Number of atoms inside the volume of an octahedral fcc($n$, $k'$)$_o$ NP, $N_{vol}(n, k')_o$, (volume count) as defined by (C.17 - 20). For the definition of $k'$ see caption of Table 3.1.

List of $N_{vol}(n, k')_o$

```
k'\n |    0      1      2      3      4      5      6      7      8      9     10     11
   0 |    1     13     55    147    309    561    923   1415   2057   2869   3871   5083
   1 |    6     38    116    260    490    826   1288   1896   2670   3630   4796   6188
   2 |   19     79    201    405    711   1139   1709   2441   3355   4471   5809   7389
   3 |   44    140    314    586    976   1504   2190   3054   4116   5396   6914   8690
   4 |   85    225    459    807   1289   1925   2735   3739   4957   6409   8115  10095
   5 |  146    338    640   1072   1654   2406   3348   4500   5882   7514   9416  11608
   6 |  231    483    861   1385   2075   2951   4033   5341   6895   8715  10821  13233
   7 |  344    664   1126   1750   2556   3564   4794   6266   8000  10016  12334  14974
   8 |  489    885   1439   2171   3101   4249   5635   7279   9201  11421  13959  16835
   9 |  670   1150   1804   2652   3714   5010   6560   8384  10502  12934  15700  18820
  10 |  891   1463   2225   3197   4399   5851   7573   9585  11907  14559  17561  20933
  11 | 1156   1828   2706   3810   5160   6776   8678  10886  13420  16300  19546  23178
```

**Table 3.3.** Number of atoms on the outermost facets of an octahedral fcc($n$, $k'$)$_o$ NP, N$_{shell}$($n$, $k'$)$_o$, as defined by (C.21 - 24). For the definition of $k'$ see caption of Table 3.1.

List of N$_{shell}$($n$, $k'$)$_o$

| $k'$\\$n$ | 0   | 1   | 2   | 3    | 4    | 5    | 6    | 7    | 8    | 9    | 10   | 11   |
|-----------|-----|-----|-----|------|------|------|------|------|------|------|------|------|
| 0         | --  | 12  | 42  | 92   | 162  | 252  | 362  | 492  | 642  | 812  | 1002 | 1212 |
| 1         | 6   | 32  | 78  | 144  | 230  | 336  | 462  | 608  | 774  | 960  | 1166 | 1392 |
| 2         | 18  | 60  | 122 | 204  | 306  | 428  | 570  | 732  | 914  | 1116 | 1338 | 1580 |
| 3         | 38  | 96  | 174 | 272  | 390  | 528  | 686  | 864  | 1062 | 1280 | 1518 | 1776 |
| 4         | 66  | 140 | 234 | 348  | 482  | 636  | 810  | 1004 | 1218 | 1452 | 1706 | 1980 |
| 5         | 102 | 192 | 302 | 432  | 582  | 752  | 942  | 1152 | 1382 | 1632 | 1902 | 2192 |
| 6         | 146 | 252 | 378 | 524  | 690  | 876  | 1082 | 1308 | 1554 | 1820 | 2106 | 2412 |
| 7         | 198 | 320 | 462 | 624  | 806  | 1008 | 1230 | 1472 | 1734 | 2016 | 2318 | 2640 |
| 8         | 258 | 396 | 554 | 732  | 930  | 1148 | 1386 | 1644 | 1922 | 2220 | 2538 | 2876 |
| 9         | 326 | 480 | 654 | 848  | 1062 | 1296 | 1550 | 1824 | 2118 | 2432 | 2766 | 3120 |
| 10        | 402 | 572 | 762 | 972  | 1202 | 1452 | 1722 | 2012 | 2322 | 2652 | 3002 | 3372 |
| 11        | 486 | 672 | 878 | 1104 | 1350 | 1616 | 1902 | 2208 | 2534 | 2880 | 3246 | 3632 |

**Table 3.4.** Corner distance, $R_c(n, k')_o$, defining the distance from the center of an octahedral fcc($n$, $k'$)$_o$ NP to its 24 ($n > 0$, $k' > 0$), 6 ($n = 0$, $k' > 0$), and 12 ($n > 0$, $k' = 0$) corners, respectively. The distances are defined by (C.28). The radii are normalized by the lattice constant $a$ of the fcc lattice. For the definition of $k'$ see caption of Table 3.1.

List of $R_c(n, k')_o / a$

| $k'$\\$n$ | 0     | 1     | 2     | 3     | 4     | 5     | 6     | 7     | 8     | 9      | 10     | 11     |
|-----------|-------|-------|-------|-------|-------|-------|-------|-------|-------|--------|--------|--------|
| 0         | 0.000 | 0.707 | 1.414 | 2.121 | 2.828 | 3.536 | 4.243 | 4.950 | 5.657 | 6.364  | 7.071  | 7.778  |
| 1         | 0.500 | 1.118 | 1.803 | 2.500 | 3.202 | 3.905 | 4.610 | 5.315 | 6.021 | 6.727  | 7.433  | 8.139  |
| 2         | 1.000 | 1.581 | 2.236 | 2.915 | 3.606 | 4.301 | 5.000 | 5.701 | 6.403 | 7.106  | 7.810  | 8.515  |
| 3         | 1.500 | 2.062 | 2.693 | 3.354 | 4.031 | 4.717 | 5.408 | 6.103 | 6.801 | 7.500  | 8.201  | 8.902  |
| 4         | 2.000 | 2.550 | 3.162 | 3.808 | 4.472 | 5.148 | 5.831 | 6.519 | 7.211 | 7.906  | 8.602  | 9.301  |
| 5         | 2.500 | 3.041 | 3.640 | 4.272 | 4.924 | 5.590 | 6.265 | 6.946 | 7.632 | 8.322  | 9.014  | 9.708  |
| 6         | 3.000 | 3.536 | 4.123 | 4.743 | 5.385 | 6.042 | 6.708 | 7.382 | 8.062 | 8.746  | 9.434  | 10.124 |
| 7         | 3.500 | 4.031 | 4.610 | 5.220 | 5.852 | 6.500 | 7.159 | 7.826 | 8.500 | 9.179  | 9.862  | 10.548 |
| 8         | 4.000 | 4.528 | 5.099 | 5.701 | 6.325 | 6.964 | 7.616 | 8.276 | 8.944 | 9.618  | 10.296 | 10.977 |
| 9         | 4.500 | 5.025 | 5.590 | 6.185 | 6.801 | 7.433 | 8.078 | 8.732 | 9.394 | 10.062 | 10.735 | 11.413 |
| 10        | 5.000 | 5.523 | 6.083 | 6.671 | 7.280 | 7.906 | 8.544 | 9.192 | 9.849 | 10.512 | 11.180 | 11.853 |
| 11        | 5.500 | 6.021 | 6.576 | 7.159 | 7.762 | 8.382 | 9.014 | 9.657 | 10.308 | 10.966 | 11.630 | 12.298 |

## S.3.3b. Cuboctahedral fcc Nanoparticles, Truncated Cubic Type

**Table 3.5.** Notation of a non-generic cuboctahedral NP fcc[N, K], fcc(n, k) for given n, k according to $2N = n + k$, $2K = 3n + k$. Structure parameters N, K describe the two generic envelope NPs, cubic fcc[N, 0), fcc[N, 1) and octahedral sc(0, K], sc(1, K] by polyhedral diameters, see Sec. C.3. In contrast, n, k characterize the deviation of the cuboctahedral NP fcc[N, M] from its generic envelope NPs.

For fcc[N, K] NPs of the cubic type, i.e. if $2N \leq K \leq 3N$, $n \geq k$, it is meaningful to use an auxiliary index n' with $n' = (n - k)/2$ resulting in a notation fcc(n', k)$_c$, see Sec. C.3, and yielding

[N, K] :   $N = n' + k$ ,   $K = 3n' + 2k$

List of [N, K]

```
k\n' |   0      1      2      3      4      5      6      7      8      9     10     11
   0 | [ 0, 0][ 1, 3][ 2, 6][ 3, 9][ 4,12][ 5,15][ 6,18][ 7,21][ 8,24][ 9,27][10,30][11,33]
   1 | [ 1, 2][ 2, 5][ 3, 8][ 4,11][ 5,14][ 6,17][ 7,20][ 8,23][ 9,26][10,29][11,32][12,35]
   2 | [ 2, 4][ 3, 7][ 4,10][ 5,13][ 6,16][ 7,19][ 8,22][ 9,25][10,28][11,31][12,34][13,37]
   3 | [ 3, 6][ 4, 9][ 5,12][ 6,15][ 7,18][ 8,21][ 9,24][10,27][11,30][12,33][13,36][14,39]
   4 | [ 4, 8][ 5,11][ 6,14][ 7,17][ 8,20][ 9,23][10,26][11,29][12,32][13,35][14,38][15,41]
   5 | [ 5,10][ 6,13][ 7,16][ 8,19][ 9,22][10,25][11,28][12,31][13,34][14,37][15,40][16,43]
   6 | [ 6,12][ 7,15][ 8,18][ 9,21][10,24][11,27][12,30][13,33][14,36][15,39][16,42][17,45]
   7 | [ 7,14][ 8,17][ 9,20][10,23][11,26][12,29][13,32][14,35][15,38][16,41][17,44][18,47]
   8 | [ 8,16][ 9,19][10,22][11,25][12,28][13,31][14,34][15,37][16,40][17,43][18,46][19,49]
   9 | [ 9,18][10,21][11,24][12,27][13,30][14,33][15,36][16,39][17,42][18,45][19,48][20,51]
  10 | [10,20][11,23][12,26][13,29][14,32][15,35][16,38][17,41][18,44][19,47][20,50][21,53]
  11 | [11,22][12,25][13,28][14,31][15,34][16,37][17,40][18,43][19,46][20,49][21,52][22,55]
```

**Table 3.6.** Number of atoms inside the volume of a cubic fcc(n', k)$_c$ NP, N$_{vol}$(n', k)$_c$, (volume count) as defined by (C.31 - 34). For the definition of n' see caption of Table 3.5.

List of N$_{vol}$(n', k)$_c$

```
k\n' |    0      1      2      3      4      5      6      7      8      9     10     11
   0 |    1     14     63    172    365    666   1099   1688   2457   3430   4631   6084
   1 |   13     62    171    364    665   1098   1687   2456   3429   4630   6083   7812
   2 |   55    164    357    658   1091   1680   2449   3422   4623   6076   7805   9834
   3 |  147    340    641   1074   1663   2432   3405   4606   6059   7788   9817  12170
   4 |  309    610   1043   1632   2401   3374   4575   6028   7757   9786  12139  14840
   5 |  561    994   1583   2352   3325   4526   5979   7708   9737  12090  14791  17864
   6 |  923   1512   2281   3254   4455   5908   7637   9666  12019  14720  17793  21262
   7 | 1415   2184   3157   4358   5811   7540   9569  11922  14623  17696  21165  25054
   8 | 2057   3030   4231   5684   7413   9442  11795  14496  17569  21038  24927  29260
   9 | 2869   4070   5523   7252   9281  11634  14335  17408  20877  24766  29099  33900
  10 | 3871   5324   7053   9082  11435  14136  17209  20678  24567  28900  33701  38994
  11 | 5083   6812   8841  11194  13895  16968  20437  24326  28659  33460  38753  44562
```

**Table 3.7.** Number of atoms on the outermost facets of a cubic fcc($n'$, $k$)$_c$ NP, $N_{shell}(n', k)_c$, (facet count) as defined by (C.35 - 38). For the definition of $n'$ see caption of Table 3.5.

List of $N_{shell}(n', k)_c$

| k\n' | 0 | 1 | 2 | 3 | 4 | 5 | 6 | 7 | 8 | 9 | 10 | 11 |
|---|---|---|---|---|---|---|---|---|---|---|---|---|
| 0 | -- | 14 | 50 | 110 | 194 | 302 | 434 | 590 | 770 | 974 | 1202 | 1454 |
| 1 | 12 | 48 | 108 | 192 | 300 | 432 | 588 | 768 | 972 | 1200 | 1452 | 1728 |
| 2 | 42 | 102 | 186 | 294 | 426 | 582 | 762 | 966 | 1194 | 1446 | 1722 | 2022 |
| 3 | 92 | 176 | 284 | 416 | 572 | 752 | 956 | 1184 | 1436 | 1712 | 2012 | 2336 |
| 4 | 162 | 270 | 402 | 558 | 738 | 942 | 1170 | 1422 | 1698 | 1998 | 2322 | 2670 |
| 5 | 252 | 384 | 540 | 720 | 924 | 1152 | 1404 | 1680 | 1980 | 2304 | 2652 | 3024 |
| 6 | 362 | 518 | 698 | 902 | 1130 | 1382 | 1658 | 1958 | 2282 | 2630 | 3002 | 3398 |
| 7 | 492 | 672 | 876 | 1104 | 1356 | 1632 | 1932 | 2256 | 2604 | 2976 | 3372 | 3792 |
| 8 | 642 | 846 | 1074 | 1326 | 1602 | 1902 | 2226 | 2574 | 2946 | 3342 | 3762 | 4206 |
| 9 | 812 | 1040 | 1292 | 1568 | 1868 | 2192 | 2540 | 2912 | 3308 | 3728 | 4172 | 4640 |
| 10 | 1002 | 1254 | 1530 | 1830 | 2154 | 2502 | 2874 | 3270 | 3690 | 4134 | 4602 | 5094 |
| 11 | 1212 | 1488 | 1788 | 2112 | 2460 | 2832 | 3228 | 3648 | 4092 | 4560 | 5052 | 5568 |

**Table 3.8.** Corner distance, $R_c(n', k)_c$, defining the distance from the center of a cubic fcc($n'$, $k$)$_c$ NP to its 24 ($n' > 0$, $k > 0$), 12 ($n' = 0$, $k > 0$), and 8 ($n' > 0$, $k = 0$) corners, respectively. The distances are defined by (C.42) and all radii are normalized by the lattice constant $a$ of the fcc lattice. For the definition of $n'$ see caption of Table 3.5.

List of $R_c(n', k)_c / a$

| k\n' | 0 | 1 | 2 | 3 | 4 | 5 | 6 | 7 | 8 | 9 | 10 | 11 |
|---|---|---|---|---|---|---|---|---|---|---|---|---|
| 0 | 0.000 | 0.866 | 1.732 | 2.598 | 3.464 | 4.330 | 5.196 | 6.062 | 6.928 | 7.794 | 8.660 | 9.526 |
| 1 | 0.707 | 1.500 | 2.345 | 3.202 | 4.062 | 4.924 | 5.788 | 6.652 | 7.517 | 8.382 | 9.247 | 10.112 |
| 2 | 1.414 | 2.179 | 3.000 | 3.841 | 4.690 | 5.545 | 6.403 | 7.263 | 8.124 | 8.986 | 9.849 | 10.712 |
| 3 | 2.121 | 2.872 | 3.674 | 4.500 | 5.339 | 6.185 | 7.036 | 7.890 | 8.746 | 9.605 | 10.464 | 11.325 |
| 4 | 2.828 | 3.571 | 4.359 | 5.172 | 6.000 | 6.837 | 7.681 | 8.529 | 9.381 | 10.235 | 11.091 | 11.948 |
| 5 | 3.536 | 4.272 | 5.050 | 5.852 | 6.671 | 7.500 | 8.337 | 9.179 | 10.025 | 10.874 | 11.726 | 12.580 |
| 6 | 4.243 | 4.975 | 5.745 | 6.538 | 7.348 | 8.170 | 9.000 | 9.836 | 10.677 | 11.522 | 12.369 | 13.219 |
| 7 | 4.950 | 5.679 | 6.442 | 7.228 | 8.031 | 8.846 | 9.670 | 10.500 | 11.336 | 12.176 | 13.019 | 13.865 |
| 8 | 5.657 | 6.384 | 7.141 | 7.921 | 8.718 | 9.526 | 10.344 | 11.169 | 12.000 | 12.835 | 13.675 | 14.517 |
| 9 | 6.364 | 7.089 | 7.842 | 8.617 | 9.407 | 10.210 | 11.023 | 11.843 | 12.669 | 13.500 | 14.335 | 15.174 |
| 10 | 7.071 | 7.794 | 8.544 | 9.314 | 10.100 | 10.897 | 11.705 | 12.520 | 13.342 | 14.169 | 15.000 | 15.835 |
| 11 | 7.778 | 8.500 | 9.247 | 10.012 | 10.794 | 11.587 | 12.390 | 13.200 | 14.018 | 14.841 | 15.668 | 16.500 |